\documentclass[aps,prb,floats,showpacs,twocolumn]{revtex4-1}
\usepackage{graphicx,color,wrapfig}
\usepackage{amsmath}
\begin{document}
\title{A Tight Binding Approach to Strain and Curvature in Monolayer Transition-Metal Dichalcogenides} 
\author{Alexander J. Pearce$^{1}$, Eros Mariani$^{2}$ and Guido Burkard$^{1}$}
\affiliation{$^{1}$Department of Physics, University of Konstanz, D-78464 Konstanz, Germany\\
$^{2}$School of Physics, University of Exeter, Stocker Rd., EX4 4QL Exeter, United Kingdom}
\begin{abstract}

We present a model of the electronic properties of monolayer transition-metal dichalcogenides based on a tight binding approach which includes the effects of strain and curvature of the crystal lattice. Mechanical deformations of the lattice offer a powerful route for tuning the electronic structure of the transition-metal dichalcogenides, as changes to bond lengths lead directly to corrections in the electronic Hamiltonian while curvature of the crystal lattice mixes the orbital structure of the electronic Bloch bands. We first present an effective low energy Hamiltonian describing the electronic properties near the K point in the Brillouin zone, then present the corrections to this Hamiltonian due to arbitrary mechanical deformations and curvature in a way which treats both effects on an equal footing. This analysis finds that local area variations of the lattice allow for tuning of the band gap and effective masses, while the application of uniaxial strain decreases the magnitude of the direct band gap at the K point. Additionally, strain induced bond length modifications create a fictitious gauge field with a coupling strength that is smaller than that seen in related materials like graphene. We also find that curvature of the lattice leads to the appearance of both an effective in-plane magnetic field which couples to spin degrees of freedom and a Rashba-like spin-orbit coupling due to broken mirror inversion symmetry. 

\end{abstract}  


\maketitle

\section{Introduction}

Monolayers of transition-metal dichalcogenides (TMDCs) are two-dimensional semiconductors and have gained great interest in recent years due to their remarkable electronic and optical properties.\cite{Strano2012} They possess a direct band gap with a magnitude in the optical range\cite{Heinz2010} and optical selection rules which allow for manipulation of the valley and spin degrees of freedom\cite{Xiao2012,Heinz2012} providing them with great potential for future device applications.

These TMDCs are a class of materials with the chemical composition MX$_2$ where M represents a transition-metal (e.g. M = Mo, W) while X corresponds to a chalcogen atom (e.g X = S, Se, Te). These atoms form a honeycomb crystal lattice analogous to graphene but possess  a semiconducting electronic structure with the band gap minima located at the K and K$'$ points of the hexagonal Brillouin zone, where the gap is due to a broken inversion symmetry. Indeed, these materials have already been shown to produce two-dimensional field effect transistors with on/off ratios orders of magnitude larger than in graphene.\cite{Kis2011} Additionally, the heavy transition-metal atoms endow the material with strong spin-orbit coupling most visibly seen in the large splitting observed in the valence band ranging from $148\,\textrm{meV}$ in MoS$_2$ to $462\,\textrm{meV}$ in WSe$_2$ crystals.\cite{Falko2015}

While there has been much interest in the optical and electronic properties of the TMDCs it has also been shown that MoS$_2$ crystal membranes have mechanical properties comparable with graphene oxide and are able withstand mechanical strains up to $10\%$.\cite{Bollinger2012} Mechanical resonator devices have been demonstrated\cite{Venstra2013} and suspended devices have been fabricated for study in optical and electronic transport experiments.\cite{Lee2013,Huang2013} These properties imply that TMDCs offer a route to studying novel nano-electromechanical systems of semiconducting membranes and the role of mechanical deformations on their electronic structure, optical and transport properties. 

In this work we theoretically investigate the role of elastic deformations on the electronic properties of the TMDCs, by incorporating the effect of strain into a tight binding model. In general, deformations of the crystal lattice lead to modifications of the bond lengths which create corrections to the electronic Hamiltonian.\cite{Ando2002,Guinea2010} In addition curvature of the crystal lattice creates corrections to the hybridization of the localized orbitals which in general lead to new couplings between Bloch bands.\cite{Ando2000,Brataas2006,Lee2011,Ochoa2013} We take an approach which allows for a description of any arbitrary lattice deformations by treating modifications of bond lengths and curvature on equal footing and including the effects of spin-orbit interactions. Then we present a low energy effective theory which describes lattice distortions, spin-orbit coupling and lattice-spin coupling. We also fit our tight binding model to first principle calculations of MoS$_2$ and use the obtained band parameters to predict the energy scales of these strain/curvature-electronic coupling mechanisms.

Due to the large spin orbit interaction and unusual spin-valley coupling within TMDCs, these materials hold particular interest for spintronics applications. Out-of-plane flexural motion of the lattice creates ripples and corrugations of the crystalline membrane which have been observed in MoS$_2$ crystals\cite{Kis2011b} and these provide electronic scattering mechanisms which are unique to two-dimensional materials.\cite{Geim2008} In light of this, lattice-spin coupling is of particular interest in spin transport as a mechanism for spin relaxation\cite{Ochoa2013} or possible spin manipulation mechanisms. To date several authors have studied spin transport and spin relaxation rates in TMDCs by symmetry based techniques,\cite{Roldan2013,Dery2013,Ochoa2014} but a precise knowledge of the coupling parameters was not clear and is a topic of this work.

The role of deformations in modifying the electronic structure has been investigated so far via density functional theory techniques\cite{Zhang2012,Zeng2012,Li2012,Shenoy2012,Walle2012,Heine2013}, which predict that strain can lead to a modification to the band gap magnitude and band effective masses. These effects have also been proposed to explain optical measurements on wrinkled MoS$_2$.\cite{Steele2013} Most dramatically it has been predicted and shown by photoluminescence experiments that strain can induce a transition between a direct and indirect band gap in monolayer MoS$_2$\cite{Shan2013,Bolotin2013,Urbaszek2013} and multilayer WSe$_2$.\cite{Javey2014} Additionally, as a consequence of the broken inversion symmetry of the crystal lattice the TMDCs have been shown to exhibit piezoelectric properties.\cite{Wang2014,Zhang2015}
 
This paper is organized as follows: In Section II we introduce the details of the tight binding model including spin-orbit interactions but without any strain or curvature. Following this, in Section III we expand the tight binding model around the high symmetry K and K$'$ points and obtain a low energy effective theory describing the conduction and valence bands. In Section IV we introduce the corrections due to mechanical deformations and curvature of the crystal lattice, then incorporate these physical effects into the low energy effective theory, and discuss the role of these new terms in the low energy theory. Finally, in Section V we present our conclusions.

\section{Model}

\begin{figure}
		\includegraphics[width=1.0\columnwidth]{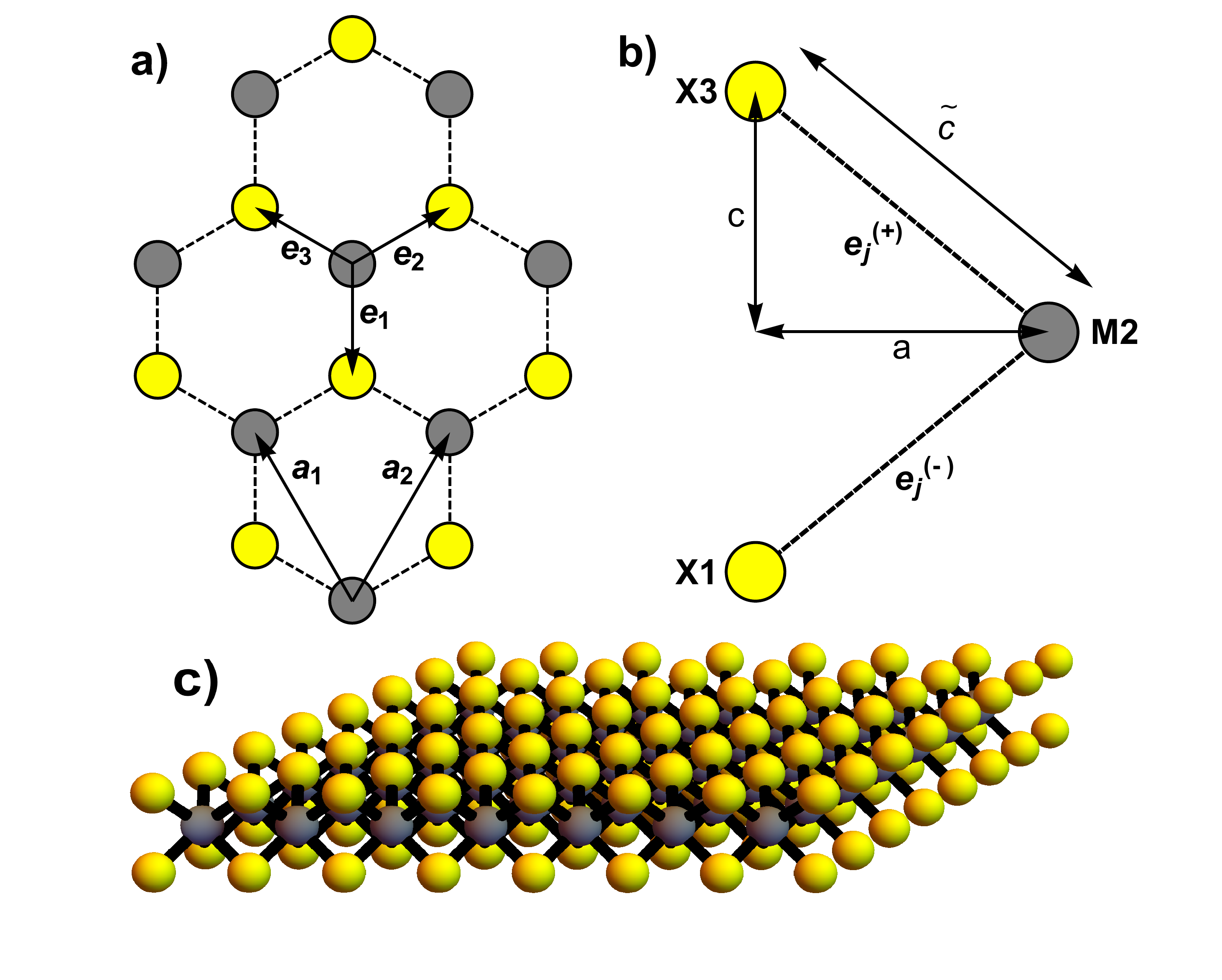}
			\caption{The crystal structure of the monolayer TMDCs. Panel a) shows a top down view of the crystal lattice, the X sites on the top and bottom layers are colored yellow while the M sites which comprise the middle layer are shown in grey. Also shown on the lattice are the nearest neighbor vectors $\mathbf{e}_j$ and the lattice vectors $\mathbf{a}_i$. Panel b) shows a side view of the lattice within one unit cell, the lattice sites X1, X3 and M2 are labeled where the X and M sites are colored yellow and grey respectively. The nearest neighbor vectors $\mathbf{e}_j^{(\pm)}$ between the M and X lattice sites are shown, and additionally the lattice parameters a, c and $\tilde{c}$ are displayed. Panel c) shows a sketch of the monolayer TMDC lattice over many unit cells.}	
\label{LatticeFig}
\end{figure}

Our calculations are based on a tight binding approach describing the multiple electronic orbitals on each lattice site that makes up the TMDC within the Koster-Slater approach.\cite{Koster1954,Harrison} The monolayer form of the TMDCs are comprised of three layers, two (top and bottom) containing only chalcogen atoms and one middle layer with only transition-metal atoms. We identify these layers by the index $l=1$ for the lower chalcogen layer, $l=2$ for the layer of transition-metal atoms, and $l=3$ for the upper chalcogen layer, a sketch of this lattice structure is shown in Fig (\ref{LatticeFig}). These layers are each separated by a distance $c=\,1.51\text{\AA}$. 

The three layers comprise a planar honeycomb lattice with the top and bottom layers arranged in a triangular lattice with the top layer lying directly above the bottom layer, while the middle layer is also made of a triangular lattice rotated by $\pi$. Together these layers comprise a honeycomb lattice constructed from two sublattices (labeled X and M), where the X sublattice is made of chalcogen atoms (labeled X1 and X3 for sites on the bottom and top layer respectively) and the M sublattice is conversely made of transition-metal atoms (labeled M2), as shown in Fig. \ref{LatticeFig}. The vectors connecting the X and M lattice sites are $\mathbf{e}_1^{(\pm)} = (0,-a,\pm c)$, $\mathbf{e}_2^{(\pm)} = \, (\sqrt{3}a/2,a/2,\pm c)$ and $\mathbf{e}_3 ^{(\pm)}= (-\sqrt{3}a/2,a/2,\pm c)$, where $a=1.84\,\text{\AA}$ and the $(\pm)$ denotes for $+$($-$) the vector connecting the M2 lattice site with upper layer X3 (lower layer X1) lattice site. As a consequence, the interatomic distance between each M atomic site and its nearest X atomic site is $\tilde{c} = \sqrt{a^2 + c^2}  = 2.37\,\text{\AA}$. A sketch of the honeycomb lattice showing these vectors is shown in Fig \ref{LatticeFig}.

It is also worth considering the next nearest neighbor vectors which connect the atomic sites of the same species and on the same layer, these are given by the six vectors $\boldsymbol{\delta}_{1} = \mathbf{e}_1^{(\pm)} - \mathbf{e}_3^{(\pm)} = a(\sqrt{3}/2,-3/2,0)$, $\boldsymbol{\delta}_{2} = \mathbf{e}_3^{(\pm)} - \mathbf{e}_1^{(\pm)} = a(-\sqrt{3}/2,3/2,0)$, $\boldsymbol{\delta}_{3} = \mathbf{e}_1^{(\pm)} - \mathbf{e}_2^{(\pm)} = a(-\sqrt{3}/2,-3/2,0)$, $\boldsymbol{\delta}_{4} = \mathbf{e}_2^{(\pm)} - \mathbf{e}_1^{(\pm)} = a(\sqrt{3}/2,3/2,0)$, $\boldsymbol{\delta}_{5} = \mathbf{e}_2^{(\pm)} - \mathbf{e}_3^{(\pm)} = a(\sqrt{3},0,0)$, and $\boldsymbol{\delta}_{6} = \mathbf{e}_3^{(\pm)} - \mathbf{e}_2^{(\pm)} = a(-\sqrt{3},0,0)$.

The electronic structure of the TMDCs is governed by the d-orbitals localized on the transition-metal atoms and the p-orbitals of the chalcogen atoms. Studies of the electronic structure of the TMDCs by DFT techniques have shown that the conduction band minimum and the valence band maximum have a strong d-orbital character, while all bands with strong p-orbital character lie at higher energies.\cite{Xiao2012,Eriksson2009,Schwingenschlogl2011,Falko2013,Guinea2013,Falko2015} To construct a realistic tight binding model we will include both the d-orbital states and higher energy p-orbitals states. Indeed, we will see later within the text that the p-orbitals are crucial for replicating the electron-hole asymmetry predicted by DFT and used to explain magneto-optical experiments\cite{Imamoglu2015,Xu2015,Ralph2015} while also giving a complete and robust model of the effects of strain. Therefore, to be fully consistent, on each M lattice site we consider 5 localized d-orbitals $d_{z^2}$, $d_{x^2-y^2}$, $d_{xy}$, $d_{xz}$ and $d_{yz}$ (we will denote $x^2-y^2$ as $x^2$ for notational convenience throughout this work), while on all X lattice sites we include the $p_x$, $p_y$ and the $p_z$ localized p-orbitals. This creates a large basis of 11 orbitals per unit cell, but we will see that we are often able to achieve a very good approximation with a smaller basis at the high symmetry points of the Brillouin zone.

The tight binding Hamiltonian in real space is given by three terms
\begin{equation} H = H_{\textrm{atom}} + H_{\textrm{hop}} + H_{\textrm{so}} \;, \label{ham} \end{equation}
where $H_{\textrm{atom}}$ describes the on-site energies of each localized orbital, $H_{\textrm{hop}}$ describes the hopping between neighbor localized orbitals, and $H_{\textrm{so}}$ describes the on-site spin-orbit interactions. We express the Hamiltonian in second quantization where the electron annihilation operator $p^{\alpha}_{\mu,i,s}$ removes an electron in a p-orbital on lattice site $\alpha=X1,X3$ in the $\mu=x,y,z$ localized orbital in the $i$-th unit cell with spin $s=\uparrow, \downarrow$, while $d_{\nu,i,s}$ removes an electron in the $\nu=z^2,xy,x^2,xz,yz$ localized d-orbital in the $i$-th unit cell with spin $s=\uparrow, \downarrow$ where all d-orbitals are localized on the M2 lattice sites. 

At this point it is convenient to change the basis to consider symmetric and antisymmetric combinations in the layer index $l=1,3$ of the localized chalcogen orbitals. This is done by introducing the the new operators $p^S_{\mu,i,s}=(p^{X1}_{\mu,i,s} + p^{X3}_{\mu,i,s})/\sqrt{2}$ and $p^A_{\mu,i,s} = (p^{X1}_{\mu,i,s} - p^{X3}_{\mu,i,s})/\sqrt{2}$. 

We will also utilize one final change of basis to the rotating orbital basis, allowing us to write the electron operators of the localized states in a form that makes their angular momentum more transparent. Therefore we introduce $d_{\pm2,i,s}=(d_{x^2,i,s} \pm i d_{xy,i,s})/\sqrt{2}$ and $d_{\pm1,i,s}=(d_{xz,i,s} \pm i d_{yz,i,s})/\sqrt{2}$ for the d-orbitals, while for the p-orbitals $p_{\pm1,i,s}^{\beta}=(p_{x,i,s}^{\beta}\pm ip_{y,i,s}^{\beta})/\sqrt{2}$ where the index $\pm2$($\pm1$) refers to the angular momentum of the rotating d(p)-orbital and for the p-orbitals states $\beta=S,A$ indicate the symmetric and antisymmetric combinations in the layer index introduced above. 

The transition-metal dichalcogenide crystal lattice is symmetric under a mirror inversion around the central layer ($z\rightarrow-z$). Due to this we categorize the eleven electronic states as even or odd under this mirror inversion. Of the states we consider the $d_{z^2,i,s}$, $d_{2,i,s}$ and $d_{-2,i,s}$ d-orbitals and the $p_{1,i,s}^{S}$, $p_{-1,i,s}^{S}$ and $p_{z,i,s}^{A}$ p-orbitals are even while the d-orbitals $d_{+1,i,s}$ and $d_{-1,i,s}$ along with the p-orbitals $p_{1,i,s}^{A}$, $p_{-1,i,s}^{A}$ and $p_{z,i,s}^{S}$ are odd under this transformation. As a consequence of this symmetry the hopping processes between the even and odd states will be zero, unless the mirror symmetry is broken somehow, say, by a perpendicular electric field or by mechanical bending of the crystal lattice. In the following we describe each term in Eq. (\ref{ham}) and its derivation in detail. 

\subsection{Atomic Hamiltonian}

The term $H_{\textrm{atom}}$ describes the on-site energies of all the localized orbitals considered within this model. The Hamiltonian is given by
\begin{align} H_{\textrm{atom}} = & \sum_{i,s} \Big[\epsilon_d^0 d^{\dagger}_{z^2,i,s}d_{z^2,i,s} + \sum_{\rho=\pm}\big(\epsilon_d^1 d^{\dagger}_{\rho1,i,s}d_{\rho1,i,s} \nonumber \\
& + \epsilon_d^2 d^{\dagger}_{\rho2,i,s}d_{\rho2,i,s}\big) + \epsilon_p^0 \big( p^{S\dagger}_{z,i,s}p^{S}_{z,i,s} + p^{A\dagger}_{z,i,s}p^{A}_{z,i,s} \big) \nonumber \\
& + \sum_{\rho=\pm}\epsilon_p^1 \big(p^{S\dagger}_{\rho1,i,s}p^{S}_{\rho1,i,s} + p^{A\dagger}_{\rho1,i,s}p^{A}_{\rho1,i,s} \big)\Big] \;, \end{align} 
where $\epsilon^{|l|}_{\lambda}$ refers to the onsite energy of the state of type $\lambda = p,d$ with angular momentum $|l|$. The numerical values of these on-site energies will be found by comparing our tight-binding model with first principle calculations, as discussed in Appendix A.2.
 
\subsection{Hopping Hamiltonian}

The term $H_{\textrm{hop}}$ describes the hopping of electron between localized orbitals on different lattice sites. Within the hopping Hamiltonian we consider three different distinct hopping processes, given by
\begin{equation} H_{\textrm{hop}} = H_{\textrm{pd}} + H_{\textrm{dd}} + H_{\textrm{pp}} \end{equation}
where $H_{\textrm{pd}}$ describes the hopping between the nearest neighbor d-orbitals on the middle layer and p-orbitals on the top and bottom layers which are connected by the vectors $\mathbf{e}^{(\pm)}_j$, $H_{\textrm{dd}}$ describes the hopping between d-orbitals located on the middle layer connected by the vectors $\boldsymbol{\delta}_j$ while $H_{\textrm{pp}}$ describes both the vertical hopping between the p-orbitals on the top and bottom layers separated by the vector $2c\hat{\mathbf{z}}$ and the in-plane hopping between p-orbitals on the same layer connected by the vectors $\boldsymbol{\delta}_j$. 

The hopping matrix element of an electron from the localized orbital $\nu'$ in the $j$-th unit cell to the $\nu$ localized orbital on the $i$-th unit cell is denoted by $t^{ij}_{\nu \nu'}$. In principle due to the large number of orbitals included in this model, we have a large number of tight binding parameters. Therefore we will employ the Koster-Slater two center approximation to express the hopping terms $t_{\nu,\nu'}^{ij}$ within the tight binding model in terms of a small number of parameters.\cite{Koster1954,Harrison} Within this scheme we are able to express all the hopping energy scales in terms of a linear combination of the parameters $V_{pp}^{\sigma}$, $V_{pp}^{\pi}$, $V_{pd}^{\sigma}$, $V_{pd}^{\pi}$, $V_{dd}^{\sigma}$, $V_{dd}^{\pi}$ and $V_{dd}^{\delta}$ where the subscript refers to the type of orbitals the hopping process is between and the superscripts $\sigma$, $\pi$ and $\delta$ refer to the bond ligands. 

Firstly, we can turn our attention to the hopping between the p and d localized orbitals within the unit cell. In general the Hamiltonian considering all nearest neighbor hopping is given by
\begin{equation} H_{\textrm{pd}} =  \sum_{\langle ij \rangle,s, \mu, \nu, \alpha} t^{ij}_{\nu\mu,\alpha} d^{\dagger}_{\nu,i,s}p^{\alpha}_{\mu,j,s} + H.c. \end{equation} 
where $t^{ij}_{\nu \mu,\alpha}$ is the hopping matrix element between the $i$-th and $j$-th unit cells and between localized orbitals of type $\nu=z^2,xy,x^2,xz,yz$ and $\mu=x,y,z$, and $\langle ... \rangle$ implies summation over all nearest neighbors. The $i$-th unit cell which is the d-orbital will be localized on a M2 lattice site, while $j$-th unit cell corresponding to a p-orbital on the $\alpha=X1,X3$ lattice site on layer 1 or 3 respectively. 

We now perform the summation over $\mu$, $\nu$ and $\alpha$ and use both changes of basis discussed earlier in the text. This procedure yields 
\begin{align} H_{\textrm{pd}} = &\; \frac{1}{\sqrt{2}} \sum_{\langle ij \rangle,s} \Big[2t_{z,z^2}^{i,j} d^{\dagger}_{z^2,i,s}p^{A}_{z,j,s} \nonumber \\
& + \sum_{\rho=\pm} \big( \sqrt{2}t_{\rho1,z^2}^{i,j} d^{\dagger}_{z^2,i,s}p^{S}_{\rho1,j,s} + \sqrt{2}t_{z,\rho1}^{i,j} d^{\dagger}_{\rho1,i,s}p^{S}_{z,j,s}\nonumber \\
& + \sqrt{2}t_{z,\rho2}^{i,j} d^{\dagger}_{\rho2,i,s}p^{A}_{z,j,s} \big) + \sum_{\rho,\rho'=\pm} \big(t_{\rho1,\rho'2}^{i,j} d^{\dagger}_{\rho'2,i,s}p^{S}_{\rho1,j,s} \nonumber \\
& + t_{\rho1,\rho'1}^{i,j} d^{\dagger}_{\rho1,i,s}p^{A}_{\rho',j,s}\big)  \Big] + H.c. \label{Hampd} \end{align} 
where the new hopping matrix element can be expressed in terms of the hopping matrix elements between the non-rotating orbitals within the Koster-Slater two center approximation. 

Next we consider the hopping between the d-orbitals within the middle layer. This Hamiltonian is given by
\begin{equation} H_{\textrm{dd}} = \sum_{\langle\langle ij \rangle\rangle,\nu,\nu',s} t_{\nu,\nu'}^{i,j} d^{\dagger}_{\nu,i,s}d_{\nu',j,s} \;,\end{equation}
where $\langle\langle...\rangle\rangle$ refers to summation over the next nearest neighbor vectors $\boldsymbol{\delta}_j$ which connect lattice sites on the same layer. We again perform the summation over the orbital index $\nu$ and use the aforementioned changes of basis to find
\begin{align} H_{\textrm{dd}} = & \frac{1}{2}\sum_{\langle\langle ij \rangle\rangle,s} \Big[2 t_{z^2,z^2}^{i,j} d^{\dagger}_{z^2,i,s}d_{z^2,j,s} + \sum_{\rho=\pm}\big(t_{\rho1,\rho1}^{i,j} d^{\dagger}_{\rho1,i,s}d_{\rho1,j,s} \nonumber \\
& + t_{\rho2,\rho2}^{i,j} d^{\dagger}_{\rho2,i,s}d_{\rho2,j,s}\big) + \Big[ \sum_{\rho=\pm}\sqrt{2}t_{z^2,\rho2}^{i,j} d^{\dagger}_{z^2,i,s}d_{\rho2,j,s} \nonumber \\
& + t_{+1,-1}^{i,j} d^{\dagger}_{+1,i,s}d_{-1,j,s} +  t_{+2,-2}^{i,j} d^{\dagger}_{+2,i,s}d_{-2,j,s} + H.c. \Big]\Big] \;, \end{align}
where the hopping matrix elements are expressed in the rotating orbital basis.

Finally we discuss the hopping between the p-orbitals, where we will consider both the in-plane hopping within the bottom and top layers of the crystal lattice and the vertical hopping between the top and bottom layer over the distance $2c$. In general the Hamiltonian which describes these processes is given by
\begin{align} H_{\textrm{pp}} = & \; H^{\textrm{in}}_{\textrm{pp}} + H^{\textrm{out}}_{\textrm{pp}} \; , \\
H^{\textrm{in}}_{\textrm{pp}} = & \sum_{\langle\langle ij\rangle\rangle,s,\mu,\mu'} t^{ij}_{\mu\mu'} \big[ p^{X1\dagger}_{\mu,i,s} p^{X1}_{\mu',j,s} + p^{X3\dagger}_{\mu,i,s} p^{X3}_{\mu',j,s} \big] \; , \\  
H^{\textrm{out}}_{\textrm{pp}} = & \sum_{i,s, \mu, \mu'} t^{ii}_{\mu \mu'} p^{X1\dagger}_{\mu,i,s}p^{X3}_{\mu',i,s} + H.c. \;. \end{align}
Under the change of basis to the symmetric and antisymmetric combinations of the rotating orbitals we find that the contribution to the in-plane hopping takes the form
\begin{equation} H^{\textrm{in}}_{\textrm{pp}} = \sum_{\langle\langle ij \rangle\rangle,s,\beta} \Big[ t^{i,j}_{z,z} p^{\beta\dagger}_{z,i,s}p^{\beta}_{z,j,s} + \frac{1}{2}\sum_{\rho,\rho'=\pm} t^{i,j}_{\rho,\rho'} p^{\beta\dagger}_{\rho1,j,s}p^{\beta}_{\rho'1,j,s}\Big]\end{equation}
where $\beta=S,A$ refers to the symmetric and anti-symmetric combinations of the localized orbitals on the top and bottom layer. Whereas the hopping Hamiltonian from the vertical hopping takes the form 
\begin{align} H^{\textrm{out}}_{\textrm{pp}} = & \sum_{i,s} \Big[ \sum_{\rho=\pm}V_{pp}^{\pi} \big( p^{S\dagger}_{\rho1,i,s}p^{S}_{\rho1,i,s} - p^{A\dagger}_{\rho1,i,s}p^{A}_{\rho1,i,s} \big)\nonumber \\
& + V^{\sigma}_{pp} \big( p^{S\dagger}_{z,i,s}p^{S}_{z,i,s} - p^{A\dagger}_{z,i,s}p^{A}_{z,i,s} \big) \Big] \;. \end{align} 
The vertical hopping therefore leads to a shift in the on-site energies of the p orbitals states. 

\subsection{Spin-orbit Interaction} 

An understanding of the role of the spin-orbit interaction is crucial for a realistic description of the electronic structure of the TMDCs. The Hamiltonian for the spin-orbit interaction in the atomic approximation is given by
\begin{equation} H_{\textrm{so}} = \frac{\hbar}{4m_e^2c^2} \frac{1}{r} \frac{dV(r)}{dr} \mathbf{L}\cdot\mathbf{S} \end{equation}
where $V({r})$  is the spherically symmetric atomic potential, $\mathbf{L}$ is the angular momentum operator and $\mathbf{S}=(s_x,s_y,s_z)$ is a vector of the spin Pauli matrices (with eigenvalues $\pm1$). We use $\mathbf{L}\cdot\mathbf{S}=(L_{+}s^{-}+L_{-}s^{+})/2+L_zs_z$ where $L_{\pm}=L_x\pm i L_y$ and $s^{\pm} = s_x\pm is_y$. We now introduce the coupling constants for the atomic spin-orbit interactions arising from both the chalcogen and transition-metal atoms, which we define to be  
\begin{align} \lambda_{X} = & \frac{\hbar}{4m_e^2c^2} \int d\mathbf{r} R^{*}_{1,l}(\mathbf{r})  \frac{1}{r} \frac{dV(r)}{dr} R_{1,l}(\mathbf{r}) \nonumber \\
\lambda_{M} = & \frac{\hbar}{4m_e^2c^2} \int d\mathbf{r} R^{*}_{2,l}(\mathbf{r})  \frac{1}{r} \frac{dV(r)}{dr} R_{2,l}(\mathbf{r}) \;,\end{align}
where $R_{1,l}(\mathbf{r})$ and $R_{2,l}(\mathbf{r})$ denote the radial atomic wave functions of the chalcogen and transition-metal atoms respectively. The precise numerical values of $\lambda_X$ and $\lambda_M$ are not fully known in the literature, but we shall take values for MoS$_2$ with $\lambda_X\simeq25\textrm{meV}$ and $\lambda_M\simeq37\textrm{meV}$.\cite{Rossier2013,Falko2015} Using these definitions we express the spin-orbit Hamiltonian in terms of the same electron operators defined in the above text as
\begin{equation} H_{\textrm{so}} = H_{\textrm{so},1} + H_{\textrm{so},2} \end{equation}
where the two terms are given by
\begin{align} H_{\textrm{so},1} = & \; \sum_{i,s,s'} \sum_{\rho=\pm} \big[ \lambda_{M}\rho \big( 2 d^{\dagger}_{\rho2,i,s} s_{z,s,s'} d_{\rho2,i,s'} \nonumber \\
& + d^{\dagger}_{\rho1,i,s} s_{z,s,s'} d_{\rho1,i,s'} \big) + \lambda_{X} \rho\big(p^{S\dagger}_{\rho1,i,s} s_{z,s,s'} p^{S}_{\rho1,i,s'} \nonumber \\
& + p^{A\dagger}_{\rho1,i,s} s_{z,s,s'} p^{A}_{\rho1,i,s'}\big)\big] \;,\\
H_{\textrm{so},2} = & \;\sum_{i,s,s'} \lambda_{M} \big( \sqrt{\frac{3}{2}} d_{z^2,i,s}^{\dagger}s^{+}_{s,s'}d_{+1,i,s'} +  \sqrt{\frac{3}{2}} d_{z^2,i,s}^{\dagger}s^{-}_{s,s'}d_{-1,i,s'} \nonumber \\
& \; + d^{\dagger}_{+2,i,s}s^{-}_{s,s'}d_{+1,i,s'} + d^{\dagger}_{-2,i,s}s^{+}_{s,s'}d_{-1,i,s'} \big) \nonumber \\
& \; + \frac{\lambda_{X}}{\sqrt{2}} \big( p^{A\dagger}_{z,i,s}s^{+}_{s,s'}p^{A}_{+1,i,s'} + p^{A\dagger}_{z,i,s}s^{-}_{s,s'}p^{A}_{-1,i,s'} \nonumber \\
& \; + p^{S\dagger}_{+1,i,s}s^{-}_{s,s'}p^{S}_{z,i,s'} + p^{S\dagger}_{-1,i,s}s^{+}_{s,s'}p^{S}_{z,i,s'} \big) +H.c. \;. \label{SOCspinflip}
\end{align}
Here the spin operator matrix elements are given by $s_{z,s,s'} = (s_z)_{s,s'}$ and $s^{\pm}_{s,s'} = (s_x\pm is_y)_{s,s'}$. This process yields two terms, a term $H_{\textrm{so},1}$ describing the band splitting in the d-bands with finite angular momentum and in the p-bands with angular momentum $l=\pm1$, and a term $H_{\textrm{so},2}$ describing the spin flip processes which couple the orbital states with even and odd symmetry. In the following sections when discussing the tight binding model without lattice deformations we will focus only on $H_{\textrm{so},1}$, as these are the dominant terms and well describe the role of spin-orbit needed to explain the experimentally observed band splittings. A discussion of the role of the spin flip terms appearing in $H_{\textrm{so},2}$ will be left until Section IV.C which analyses these and other terms coupling the odd and even sectors of the Hamiltonian.

\section{Tight Binding Model Near the K Point}

In this section we focus our attention to the band structure around the K points in Brillouin zone, which correspond to the locations of the band gap minima. There are two inequivalent K points within the Brillouin zone, labeled K and K$'$. The K points are located at $\mathbf{k}_D = \tau(4\pi/3\sqrt{3}a,0)$, where we have introduced the valley index $\tau=\pm1$ with $\tau=1$\;($\tau=-1$) referring to the K(K$'$) point. 

To study the tight binding model in the vicinity of the K points we transform the electron operators to the k-space representation using  
\begin{equation} p^{\beta}_{\mu,i,s} = \frac{1}{\sqrt{N}} \sum_\mathbf{k} e^{i\mathbf{k}\cdot\mathbf{R}_i} p^{\beta}_{\mu,\mathbf{k},s} \end{equation} and 
\begin{equation} d_{\nu,i,s} = \frac{1}{\sqrt{N}} \sum_\mathbf{k} e^{i\mathbf{k}\cdot\mathbf{R}_i} d_{\nu,\mathbf{k},s} \end{equation}
where $\beta=S,A$, $\mu = \pm1,z$ and $\nu = \pm1,\pm2,z^2$ and $\mathbf{R}_i$ is the i-th unit cell in real space. This process yields the electronic Hamiltonian 
\begin{equation} H = \sum_{\mathbf{k}} \psi^{\dagger}_{\mathbf{k}} H_{\mathbf{k}} \psi_{\mathbf{k}} \;. \end{equation}
Here $H$ denotes the Hamiltonian describing the electronic Bloch bands in the space of
\begin{equation} \psi_{\mathbf{k}} = \big( \psi_{\mathbf{k}}^E , \psi_{\mathbf{k}}^O \big) \;, \end{equation}
where the basis is split into its even and odd subsectors, which are individually given by 
\begin{align} \psi_{\mathbf{k}}^E = &\, \big( d_{z^2,\mathbf{k},s}, d_{\tau2,\mathbf{k},s}, d_{-\tau2,\mathbf{k},s}, p^{S}_{\tau1,\mathbf{k},s}, p^{S}_{-\tau1,\mathbf{k},s}, p^{A}_{z,\mathbf{k},s} \big)^T \nonumber \\
\psi_{\mathbf{k}}^O = &\, \big(d_{\tau1,\mathbf{k},s}, d_{-\tau1,\mathbf{k},s}, p^{A}_{\tau1,\mathbf{k},s}, p^{A}_{-\tau1,\mathbf{k},s}, p^{S}_{z,\mathbf{k},s} \big)^T \;.\label{basis}\end{align}
We note that this basis will be used near the K points and at these high symmetry points the basis is most succinctly defined with the valley index $\tau$ in mind, therefore we see that the interchange of valleys also interchanges the order of some of the orbital states within the basis. We now consider the full Hamiltonian discussed so far, with the exception of $H_{\textrm{so},2}$ which we will introduce in Section IV.C. This Hamiltonian is block diagonal in the subsectors with even and odd symmetries with respect to mirror inversion around the central layer, allowing the Hamiltonian to be expressed as
\begin{equation} H_{\mathbf{k}} = \left( {\begin{array}{cccc}
H_{\mathbf{k},dd}^E & H_{\mathbf{k},pd}^E & 0 & 0 \\
H_{\mathbf{k},pd}^{E\dagger} & H_{\mathbf{k},pp}^E & 0 & 0 \\
0 & 0 & H_{\mathbf{k},dd}^O & H_{\mathbf{k},pd}^O \\
0 & 0 & H_{\mathbf{k},pd}^{O\dagger} & H_{\mathbf{k},pp}^O \\
\end{array}  } \right) \;, \label{Fullham}  \end{equation}
where the odd subblock lives within the space of the $\psi_{\mathbf{k}}^O$ while the even subblock lives within the space of $\psi_{\mathbf{k}}^E$, as given in Eq. (\ref{basis}). Around the K point DFT calculations have shown that the lowest conduction and highest valence bands all prossess an even symmetry, therefore here we will present the even subblock of the Hamiltonian and the odd subblock is presented in full in Appendix A.3. The odd states will play an important role upon considering the effects of curvature and spin flip processes arising from the spin-orbit interaction which we will address in due course in Sections IV.B and IV.C. 

We now find the electronic dispersion around the K points by expanding within the small momentum $\mathbf{q}$, where $|\mathbf{q}|a \ll 1$ using $\mathbf{k} = \mathbf{k}_D + \mathbf{q}$  with the two Dirac points $\mathbf{k}_D = \tau(4\pi/3\sqrt{3}a,0)$. Performing this expansion we find the Hamiltonian to be given by the subblocks  
\begin{align} H_{\mathbf{q},dd}^{E,\tau} = &\, \left( {\begin{array}{ccc}
V_{D0}(\mathbf{P}) & v_{dd}^{(0)}\mathbf{Q} & v_{dd}^{(0)}\mathbf{Q}^{\dagger} \\
v_{dd}^{(0)}\mathbf{Q}^{\dagger} & V^{(+)}_{D2}(\mathbf{P}) & v_{dd}^{(2)}\mathbf{Q}^{\dagger} \\
v_{dd}^{(0)}\mathbf{Q} & v_{dd}^{(2)}\mathbf{Q} & V^{(-)}_{D2}(\mathbf{P}) \\
\end{array}  } \right) 
\label{HamKpointdd} \;, \\
H_{\mathbf{q},pp}^{E,\tau} = &\, \left( {\begin{array}{ccc}
V^{(+)}_{P1}(\mathbf{P}) & -v_{pp}\mathbf{Q} & 0 \\
-v_{pp}\mathbf{Q}^{\dagger} & V^{(-)}_{P1}(\mathbf{P}) & 0 \\
0 & 0 & V_{P0}(\mathbf{P}) \\
\end{array}  } \right) \;,
\label{HamKpointpp} \\
H_{\mathbf{q},pd}^{E,\tau} = &\, \left( {\begin{array}{ccc} 
i \tau v_{pd}^{(1)}\mathbf{Q} & -i \tau K_{pd}^{(1)}\mathbf{P} & v_{pd}^{(0)} \mathbf{Q}^{\dagger} \\
i\tau K_{pd}^{(2s)}\mathbf{P} & -i \tau v_{pd}^{(2a)}\mathbf{Q}^{\dagger} & -v_{pd}^{(2)}\mathbf{Q} \\
i \tau v_{pd}^{(2a)}\mathbf{Q}^{\dagger} & -i \tau v_{pd}^{(2s)}\mathbf{Q}^{} & -K_{pd}^{(2)}\mathbf{P} \\
\end{array}  } \right) \,,
\label{HamKpointpd} \end{align}
where $\mathbf{Q} = \mathbf{q} + (a/4)\mathbf{q}^{\dagger 2}$ and $\mathbf{P}=1-(3a^2/4)|\mathbf{q}|^2$ with $\mathbf{q} = \tau q_x + i q_y$. Here we have expanded around the K points to second order in $\mathbf{q}$, as was found to be a requirement to obtain a good fitting with the DFT model. The fitting procedure is discussed in more detail in Appendix A.2. 

Precisely at the two K points (where $\mathbf{q}=0$) each d-orbital couples only with one of the p-orbitals, these pairings described by our tight-binding agree with high precision with other combined DFT and tight-binding models\cite{Guinea2013,Fang2015,Ridolfi2015} and symmetry based approaches.\cite{Falko2013} The full expressions for the energies $K^{(i)}_{pd}$ (with $i=1,2,2s$) and $V_j$ (with $j=P0,P1,D0,D2$) and the group velocities $v_{pd}^{(k)}$ (with $k=0,1,2,2a,2s$), $v_{dd}^{(n)}$ (with $n=0,2$) and $v_{pp}$ are presented in Appendix A.1. We then fit this tight-binding model to first principle calculations for the electronic structure of MoS$_2$ to determine the Koster-Slater parameters, we describe this process in detail and present the values in Appendix A.2.   

\subsection{Low Energy Effective Theory}

Now we aim to construct the effective theory of the lowest energy d-bands around the K point from the Hamiltonian presented in Eq. (\ref{Fullham}). Firstly we identify the high energy components of the model as the on-site energies of the p-orbital state $V_{P1}$, and the energy $K_{pd}^{(2)}$ which splits off the bands with $p^A_z$ and $d_{-\tau2}$ orbital characters to energies with large magnitudes. This allows us to project the 6 band even sector Hamiltonian onto the 2 d-bands in the space of $(d^{\tau}_{z^2,\mathbf{q},s}, d^{\tau}_{\tau2,\mathbf{q},s})$ via a Schrieffer-Wolf transformation.\cite{Schrieffer1966,Winker} Therefore we consider only the even subsector of the Hamiltonian in Eq. (\ref{Fullham}) which we write in a block form identifying the high and low energy sector, this allows us to write the even sector in the form
\begin{equation} H_{\mathbf{q},s}^{E,\tau} = \left( {\begin{array}{cc}
H_{\textrm{Low}} & V \\
V^{\dagger} & H_{\textrm{High}} \\
\end{array}  } \right) \;.  \end{equation} 
The high energy sector of the Hamiltonian $H_{\textrm{High}}$ is given by the $4\times4$ block in the space $(d^{\tau}_{\mathbf{q},-\tau2,s}, p^{S,\tau}_{\mathbf{q},\tau,s}, p^{S,\tau}_{\mathbf{q},-\tau,s}, p^{A}_{\mathbf{k},z,s})$. To contrast, $V$ consists of the $2\times4$ block which couples the low and high energy sub blocks, with matrix elements which are small with respect to the difference in energy eigenvalues of the high and low energy sectors of the Hamiltonian, satisfying $|\epsilon_{\textrm{Low}}-\epsilon_{\textrm{High}}|\gg|V|$. Then following the Schrieffer-Wolf transformation procedure we decouple the high and low energy sub blocks up to second order in matrix elements of V. This process yields 
\begin{align} H_{\mathbf{q},\textrm{eff}}^{E,\tau} = & H_{\textrm{0}} + H_{\textrm{trig}} + H_{\textrm{cub}}  + H_{\textrm{so}}\;,\\
H_{\textrm{0}} = & \left( {\begin{array}{cc}
\Delta_c & v\mathbf{q}  \\
v \mathbf{q}^{\dagger}  & \Delta_v  \\
\end{array} } \right) + \left( {\begin{array}{cc}
\beta |\mathbf{q}|^2 & 0 \\
0 & \alpha |\mathbf{q}|^2 \\
\end{array} } \right) \;, \label{Ham2by20} \\
H_{\textrm{trig}} = &\kappa \left( {\begin{array}{cc}
0 & \mathbf{q}^{\dagger 2} \\
 \mathbf{q}^2 & 0 \\
\end{array} } \right) \;,  \label{Ham2by2trig}  \\
H_{\textrm{cub}} = & |\mathbf{q}|^2 \left( {\begin{array}{cc}
\tau\omega_c|\mathbf{q}|\cos(3\varphi_{\mathbf{q}}) &\omega \mathbf{q} \\
\omega \mathbf{q}^{\dagger} & \tau\omega_v|\mathbf{q}|\cos(3\varphi_{\mathbf{q}}) \\
\end{array} } \right) \;, \\ 
H_{\textrm{so}} = & \left( {\begin{array}{cc}
0 & 0 \\
0 & 2 \tau \lambda_{M} s_{z} \\
\end{array} } \right) \;.\label{HamSoc} \end{align} 
where the parameters $\Delta_c$, $\Delta_v$, $v$, $\alpha$, $\beta$, $\omega$, $\omega_c$, $\omega_v$ and $\kappa$ are expressed in terms of the parameters of the six band tight-binding model and $\varphi_{\mathbf{q}} = \arctan(q_y/q_x)$.

This is the final result for the low energy effective theory, firstly it gives the same terms as the Hamiltonian proposed by Xiao {\it et al}\cite{Xiao2012} while also including terms which are higher order in $q$, like electron-hole asymmetry, trigonal warping and cubic terms which have been proposed by other authors.\cite{Falko2015,Asgari2013,Falko2013,Fang2015,Liu2013} Here we systematically  include all terms up to second order in momentum, but for completeness also include cubic corrections which at the large momenta $q\geq 0.2/a $ become important for the correct description of the electronic bands. This low energy effective theory based on tight-binding approach agrees well with Hamiltonians derived from $\mathbf{k}\cdot\mathbf{p}$ methods.\cite{Falko2015,Falko2013}  

Comparing the expressions found for the six band model and those found for the parameters of the two band low energy effective theory we determine the numerical values for the case of MoS$_2$. We find for the band edges $\Delta_c = 1.78\textrm{eV}$ and $\Delta_v = -0.19\textrm{eV}$, the group velocities of the Dirac like term $v=2.44\textrm{eV\AA}$, the effective masses $\beta=0.21\textrm{eV\AA$^{2}$}$, $\alpha=0.71\textrm{eV\AA$^{2}$}$, the higher order in momentum trigonal corrections $\kappa=0.32\textrm{eV\AA$^{2}$}$ and cubic corrections $\omega=-0.56\textrm{eV\AA$^{3}$}$, $\omega_c=-0.67\textrm{eV\AA$^{3}$}$ and $\omega_v=1.68\textrm{eV\AA$^{3}$}$. These values show reasonable agreement with those found in the recent review by Korm\'{a}nyos \textit{et al}\cite{Falko2015}, and the small discrepancies between the numerical values of our parameters and those reported from fitting to $\mathbf{k}\cdot\mathbf{p}$ model arise due to our fitting to the six band model in contrast to the two band model. 

There have been several theoretical predictions of a small spin splitting in the conduction band of the order of $\simeq3\,\textrm{meV}$ in MoS$_2$ and $\simeq-30\,\textrm{meV}$ in WS$_2$.\cite{Rossier2013,Burkard2014,Falko2015} The appearance of this spin splitting in the $d_{z^2}$ band, with $l=0$, can be understood due to its mixing with $p^{S}_{-\tau1}$ chalcogen orbitals leading to finite orbital angular momentum. Within our low energy effective theory we find sub leading corrections with the form $\tau \tilde{\lambda}_X s_{z} d^{\dagger}_{\mathbf{q},z^2,s} d_{\mathbf{q},z^2,s}$ which account for a spin splitting in the conduction band. Evaluating $\tilde{\lambda}_X$ for the parameters of MoS$_2$ we find $\simeq10\,\textrm{meV}$, but crucially this calculation was restricted to the space of the even orbital states and therefore ignores the contribution from second order processes which will be of the same order as this small energy scale $\tilde{\lambda}_X$. In Section IV.C we will address the contributions due to second order processes to the term $H_{\textrm{so}}$.

\section{Analysis of Strain}

In this section we will introduce the effect of lattice deformations of the electronic structure of the TMDCs. The effects of strain are included within the tight-binding model by allowing for the modifications of the bond length between atomic sites of the crystal lattice and the relative orientation of the localized electronic orbitals. 

Bending of the TMDC breaks the mirror symmetry under inversion around the central layer and will therefore create terms in the electronic Hamiltonian which couple the even and odd sectors. To fully account for this coupling of odd symmetry electronic states with even states we introduce mechanical deformation into the electronic Hamiltonian in two steps. In Section IV.A we first investigate the role of mechanical deformation which leads to modifications of the bond lengths of the TMDC crystal lattice within the even sector of the Hamiltonian presented above. Secondly, in Section IV.B we consider curvature of the two-dimensional TMDC and present the corrections to the even electronic states due to coupling to the odd sector of the Hamiltonian. Finally in Section IV.C we will incorporate both of these corrections into the low energy effective theory already presented in the previous section.   

\subsection{Mechanical Deformations} 

In this section we study the effects of mechanical deformations on the even sector of the TMDC electronic Hamiltonian. We work within the framework of continuum elasticity theory and describe any in-plane deformation by a two dimensional vector field $\mathbf{u}(\mathbf{r})$, while out-of-plane deformations are described by a scalar field $h(\mathbf{r})$. A generic atom at position $\mathbf{r}$ is thus shifted to $\mathbf{r}' = \mathbf{r} + \mathbf{u}(\mathbf{r}) + \hat{\mathbf{z}} h(\mathbf{r})$. In this work we assume that the three layers which make up the TMDC lattice move uniformly and neglect the effects due to shear interlayer motion. 

The lattice deformations lead to displacements of the positions of the atomic sites which change the bond lengths between atoms. This can be viewed as a modification of the hopping matrix elements, where the matrix elements $t_{\mu \mu'}^{ij}$ between a localized orbital $\mu$ in unit cell $i$ and a localized orbital $\mu'$ in the unit cell $j$ is transformed as  
\begin{equation} t_{\mu \mu'}^{ij} \rightarrow t_{\mu \mu'}^{ij} + \delta t_{\mu \mu'}^{ij} \simeq t_{\mu \mu'}^{ij} + \frac{\partial t_{\mu \mu'}^{ij}}{\partial l_0} \delta l^{ij}_{\mu\mu'} \label{StrainTrans} \end{equation} 
where $\delta t_{\mu \mu'}^{ij}$ is the correction due to the change of the bond length, $l_0$ is the undeformed bond length and $\delta l^{ij}_{\mu\mu'}$ is the modification of the bond length given by $\delta l^{ij}_{\mu\mu'} = l^{ij}_{\mu\mu'} - l_0$. The derivative can be estimated as $\partial t_{\mu \mu'}^{ij}/\partial l_0 = -\Gamma_{\mu \mu'} (t_{\mu \mu'}^{ij}/ l_0)$ where $\Gamma_{\mu \mu'} = \partial \ln t_{\mu \mu'}^{ij}/\partial \ln l_0$ is the electron Gr\"{u}neisen parameter. To the best of our knowledge the electron Gr\"{u}neisen parameters for the TMDC family of materials are currently not known in the literature, but can be obtained from a careful comparison of first principle calculations of the electronic structure of strained TMDCs and the model presented in this work or by estimation of the electron-phonon couplings measured in experiments. Typically the electron Gr\"{u}neisen parameter is of the order of one, therefore in the following for numerical evaluation we will take all $\Gamma_{\mu \mu'}$ to be unity.   

We now extend the model presented in Section III to include lattice deformations utilizing the transformation described in Eq. (\ref{StrainTrans}). In the even sector of the Hamiltonian we obtain the correction 
\begin{equation} \delta H^{\tau}_{} = \left( {\begin{array}{cc}
\delta H_{dd}^{E,\tau} & \delta H_{pd}^{E,\tau} \\
\delta H_{pd}^{E,\tau\dagger} & H_{pp}^{E,\tau} \\
\end{array}  } \right) \;, \label{FullhamStrain}  \end{equation}
where the sub blocks are given by
\begin{align} \delta H_{dd}^{E,\tau} = &\, \left( {\begin{array}{ccc}
D_1 & F^{(\tau)}_4 & F^{(\tau)\dagger}_4 \\
F^{(\tau)\dagger}_4 & D_2 &  F^{(\tau)\dagger}_3 \\
F^{(\tau)}_4 & F^{(\tau)}_3 & D_2 \\
\end{array}  } \right) 
\label{HamKpointStraindd} \;, \\
\delta H_{pd}^{E,\tau} = &\, \left( {\begin{array}{ccc} 
i \tau F^{(\tau)}_5& - 2 i \tau D_5 & F^{(\tau)\dagger}_6 \\
2 i\tau D_7& -i\tau F^{(\tau)\dagger}_9 & -F^{(\tau)}_8 \\
i\tau F^{(\tau)\dagger}_9 & -i \tau F^{(\tau)}_7 & -2 D_8 \\
\end{array}  } \right) \;, \\
\delta H_{pp}^{E,\tau} = &\, \left( {\begin{array}{ccc} 
D_{10} & -F^{(\tau)}_{12} & 0 \\
-F^{(\tau)\dagger}_{12} & D_{10} & 0 \\
0 & 0 & D_{11} \\
\end{array}  } \right) \;,
\label{HamKpointStrainpd} \end{align}
and 
\begin{align} D_k = &\; \gamma_k\textrm{Tr}[\varepsilon_{ij}]\;, \\
F^{(\tau)}_k = &\; \gamma_k\big(\varepsilon_{yy} - \varepsilon_{xx} + i\tau 2 \varepsilon_{xy}\Big)\;. \end{align}
Here we used the strain tensor for a two-dimensional membrane given by\cite{Landaubook}
\begin{equation} \varepsilon_{ij} = \frac{1}{2}\big(\partial_i u_j + \partial_j u_i + ( \partial_i h )( \partial_j h) \big)\;. \end{equation}
The couplings $\gamma_k$ are composed of the electron Gr\"{u}neisen parameters, Koster-Slater parameters, and the crystal lattice parameters and are given explicitly in Appendix B.I. 

We find that the corrections to the electronic momenta have the same form as those which appear in monolayer graphene.\cite{Guinea2010} They are linear in the in-plane deformations which cause variations of bond length and quadratic in out-of-plane deformations which lead to variations in bond length. This quadratic coupling arises due to the symmetry of the lattice with respect to the $x-y$ plane. As a consequence the corrections to intralayer ($F^{(\tau)}_{k}$ with $k=3,4,12$) and interlayer ($F^{(\tau)}_{k}$ with $k=5,..,9$) hopping terms possess the same form regardless of the different orientations. From a more microscopic understanding we find that due to our choice of basis all linear out-of-plane terms which arise in the corrections to the interlayer hopping terms cancel. This is in contrast to corrections to interlayer hopping terms which arise in some other two-dimensional materials such as bilayer graphene.\cite{Oppen2012} We also observe corrections to the energies which appear in the unstrained Hamiltonian which are only sensitive to local dilations in the size of the lattice $D_k$ and are unaffected by shear deformations. Due to the broken inversion symmetry of the crystal structure the $D_k$ terms take different coefficients allowing for modulation of the magnitude of band gaps, in contrast to inversion-symmetric two dimensional materials such as graphene. 


The terms $F^{(\tau)}_k$ behave as fictitious gauge fields acting as corrections to the electronic momenta in the unperturbed six band Hamiltonian presented in Section II. These fictitious gauge fields are time reversal invariant, where time reversal symmetry exchanges the two valleys within the Brillouin zone, with $H^{E,\tau=+}_{\mathbf{q}} + \delta H^{\tau=+}  = \big(H^{E,\tau=-}_{-\mathbf{q}}\big)^{*} + \big(\delta H^{\tau=-}\big)^{*}$. There have been proposals to use fictitious gauge fields to create pseudo-magnetic fields in MoS$_2$ devices\cite{Guinea2014} and the analogous phenomenon has been observed in strained graphene nanobubbles.\cite{Crommie2010} As a consequence of the time reversal invariant nature of the fictitious gauge fields the pseudo-magnetic fields will occur with opposite signs in each valley, as we expect as elastic deformations cannot break time reversal symmetry.

This description of strain we present here only treats electronic states near the K points of the Brillouin zone. It has been shown theoretically and experimentally that strain of the crystal lattice shifts the position of the band edge of the $\Gamma$ point, such that it becomes important for providing a description of the electronic dispersion near the band edges at $\delta L/L \approx 5\%$,\cite{Shenoy2012,Walle2012,Heine2013} where $L$ is the unstrained sample size and $\delta L$ the modification of the sample size under strain. Therefore our low energy theory is only valid to strains of order $\varepsilon_{ij}\leq0.05$, and at larger strains the model would require an extension to describe the $\Gamma$ and Q points in addition to the K point.      

\subsection{Curvature}

\begin{figure}[ht!]
	\centering
		\includegraphics[width=1.0\columnwidth]{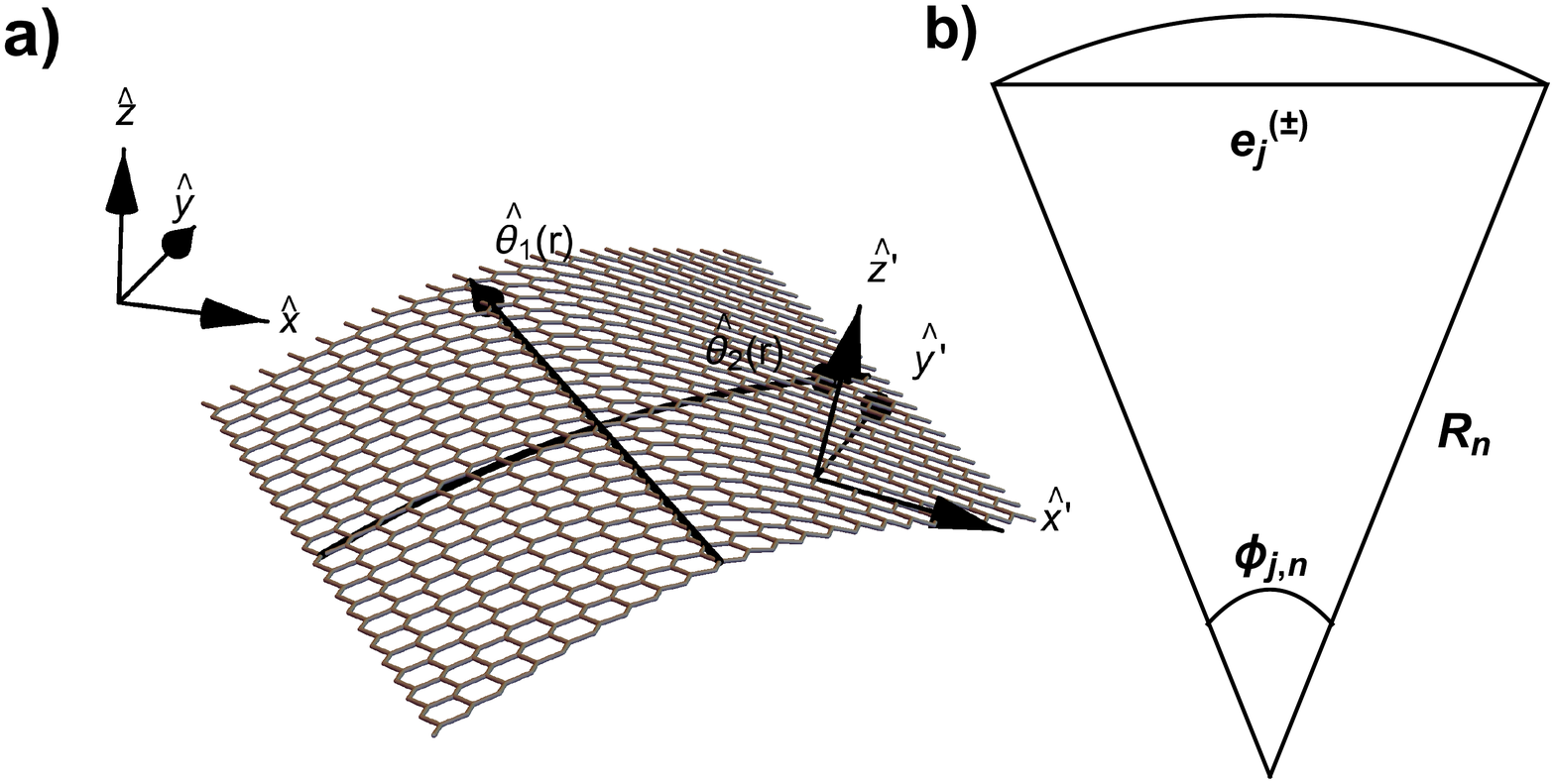}
			\caption{Panel a) shows a sketch of the curved surface of a TMDC, shown in the diagram is the both the global Cartesian coordinate system and the local Cartesian coordinate system on the curved surface. Also shown are the vectors $\hat{\theta}_1(\mathbf{r})$ and $\hat{\theta}_2(\mathbf{r})$ pointing along the principal curvatures formed by curvature of the TMDC sheet.	 Panel b) shows a cross section of the curved TMDC sheet labelling the angle $\phi_{j,n}$, the nearest neighbor bond length $\mathbf{e}_j^{(\pm)}$ and the radius of the cylinder $R_{n}$.}\label{curvatureFig}
\end{figure}

Due to curvature of the two-dimensional crystal lattice of the TMDC, hopping between localized electronic states which are forbidden by symmetry in the flat configuration become allowed, and this creates a coupling of the odd and even sectors of the electronic Hamiltonian presented in Section III. An arbitrarily curved two-dimensional surface can be described by its two principal curvatures $\kappa_1(\mathbf{r})$ and $\kappa_2(\mathbf{r})$ which we define locally. These local curvatures are expressed along principal directions at angles $\theta_1(\mathbf{r})$ and $\theta_2(\mathbf{r})$ defined from the global x-axis respectively. A sketch showing these definitions can be seen in Fig. \ref{curvatureFig}.  

To allow us to describe the role of curvature on the same footing as the mechanical deformations discussed in the last section we consider the principal curvatures in the Monge parameterization and we express the curvature in terms of the scalar height field $h(\mathbf{r})$ that we have already introduced. The two principal curvatures are the eigenvalues of the shape operator $\mathcal{S}=\mathcal{F}_1^{-1}\mathcal{F}_2$ where $\mathcal{F}_1$ is the first fundamental form and $\mathcal{F}_2$ is the second fundamental form.\cite{diffgeobook} Additionally the corresponding eigenvectors of $\mathcal{S}$ give the principal directions and allow for the determination of $\theta_n(\mathbf{r})$. We assume that we can only consider smooth curvatures, such that $\partial_i h\partial_j h \ll 1$ with $i,j\in \{x,y\}$ and write the first and second fundamental forms of a two dimensional surface as 
\begin{align} \mathcal{F}_1 =& \left( {\begin{array}{cc}
 1 + (\partial_x h)^2 & \partial_x h\partial_y h \\
 \partial_y h\partial_x h & 1 + (\partial_y h)^2 \\
\end{array}  } \right) \;, \\
\mathcal{F}_2 =& \left( {\begin{array}{cc}
 \partial_x^2 h & \partial_x\partial_y h \\
\partial_y\partial_x h & \partial_y^2 h \\
\end{array}  } \right) \;.
\end{align}
Here the first fundamental form $\mathcal{F}_1$ represents the metric of the surface and the second fundamental form $\mathcal{F}_2$ corresponds to the curvature tensor of the two-dimensional membrane. This process yields the eigenvalues $\kappa_{n}(\mathbf{r}) = (\partial_x^2 h + \partial_y^2 h + (-1)^{n} [(\partial_x^2 h - \partial_y^2 h)^2 + 4(\partial_x\partial_yh)^2]^{\frac{1}{2}})/2$ where $n=1,2$. 

To introduce these effects into the tight binding model we assume that the curvature along each principal direction can be modeled as a cylinder, where the principal curvature is given by $\kappa_n(\mathbf{r})=R^{-1}_n(\mathbf{r})$ with $R_n(\mathbf{r})$ being the local radius of the cylinder. We assume only smooth curvatures and corrugations such that $a/R_n(\mathbf{r})\ll1$ and that it will be sufficient to consider terms to first order in $a R^{-1}_n(\mathbf{r})\ll1$ to capture all the essential physics. This means that the mutual curvature effects between the two principal curvatures are negligible giving only contributions at higher orders in $a/R_n(\mathbf{r})$ and we may model the two principal curvatures independently. The two principal directions are related by $\hat{\theta}_2(\mathbf{r}) = \hat{z}'(\mathbf{r}) \times \hat{\theta}_1(\mathbf{r})$ where $\hat{z}'(\mathbf{r})$ is the vector normal to the curved surface at position $\mathbf{r}$, however for smooth curvatures and corrugations $\hat{z}'(\mathbf{r})$ maybe replaced by $\hat{z}(\mathbf{r})$. The two angles which define the two principal directions are related by $\theta_2(\mathbf{r}) = \theta_1(\mathbf{r}) \pm \pi/2$ as they are orthogonal.

Since we assume that along each principal direction the curvature is modeled as a cylinder, we proceed in the spirit of calculations studying curvature effects in carbon nanotubes.\cite{Ando2000,Brataas2006} To account for curvature within the Koster-Slater approximation we must consider the shift in the bond directions, under a cylindrical curvature along the principal direction $\theta_n(\mathbf{r})$. The corrections to the bond directions can be obtained in terms of the angle $\phi_{j,n}$ (a sketch showing a definition of the angle can be seen in Fig. \ref{curvatureFig}.b). The angle $\phi_{j,n}$ can then be expanded in the small parameter $a/R_{n}$ yielding  
\begin{equation} \phi_{j,n} \simeq \frac{1}{R_{n}}|e_{x,j}^{(\pm)}\cos\theta_n(\mathbf{r}) + e_{y,j}^{(\pm)}\sin\theta_n(\mathbf{r})| \;. \end{equation}
Where $e_{\alpha,j}^{(\pm)}$ is the $\alpha\in \{x,y\}$ component of the nearest neighbor vector for the j-th bond length. An analogous expression can also be found for the next nearest neighbor hopping between lattice sites on the layer. 

As we are interested in the physics near the K and K$'$ points we consider only the terms arising due to the momenta precisely at the K points, with $\mathbf{q}=0$. This approximation of neglecting the small momentum deviations around the K points disregards terms which act to renormalize the group velocities, an effect that turns out to be small. Introducing these corrections into the Koster-Slater approximation it is straight forward to calculate the Hamiltonian which couples the even and odd states under curvature. 
  
\begin{equation} \delta H^{\tau}_{\textrm{curv}} = \left( {\begin{array}{cc}
 \delta H_{dd,\textrm{curv}} & \delta H_{dp,\textrm{curv}} \\
 \delta H_{pd,\textrm{curv}} & \delta H_{pp,\textrm{curv}} \\
\end{array}  } \right) \;,\label{HamCurva} \end{equation}
where the subblocks are given by 
\begin{align} \delta H_{dd,\textrm{curv}} = & \frac{a}{\tilde{R}}\left( {\begin{array}{cc}
- i \tau V_1 e^{i\tau2\theta} & - i \tau V_1 e^{-i\tau2\theta}  \\
- i\tau V_2 e^{-i\tau2\theta} & i\tau V_3  \\
i\tau V_3 & -i \tau V_2 e^{i\tau2\theta}  \\
\end{array}  } \right) \;, \\
\delta H_{dp,\textrm{curv}} = & \frac{a}{\tilde{R}}\left( {\begin{array}{ccc}
 - i\tau V_4 e^{-i\tau2\theta} & i\tau 2V_4 & -\tau V_5 e^{i\tau2\theta} \\
 i\tau 2V_6 & i\tau V_7 e^{i\tau2\theta} & V_8 e^{-i\tau2\theta} \\
 -i\tau V_7 e^{i\tau2\theta} & i\tau V_6 e^{-i\tau2\theta} & 2V_8 \\
\end{array}  } \right) \;, \\
\delta H_{pd,\textrm{curv}} = & \frac{a}{\tilde{R}}\left( {\begin{array}{cc}
 - V_9 e^{i\tau2\theta} & - 2\sqrt{2}V_{8} \\
 \sqrt{2}V_{8} e^{-i\tau2\theta} & - V_9 e^{i\tau2\theta} \\
 -\tau 2V_{10} & - \tau V_{10} e^{-i\tau2\theta} \\
\end{array}  } \right) \;, \end{align}
where $\tilde{R}^{-1} = R_1^{-1} + R_2^{-1}$. The prefactors $V_j$ (with $j=1,..,10$) are all energy scales made up from Koster-Slater parameters and lattice parameters and are given in full detail in Appendix B.2. 

We note that the odd and even symmetry p-orbitals only produce curvature induced matrix elements which couple the even and odd sectors of the Hamiltonian at second order in $a/\tilde{R}$. 

\subsection{Low Energy Effective Theory}

With the inclusion of all effects that we have thus far covered within this work, the tight binding model for unstrained TMDCs and the effects of spin-orbit coupling, mechanical deformations and curvature, we obtain a model containing a large quantity of information. In this section we aim to produce a simplified low-energy effective theory describing the conduction and valence bands near the K and K$'$ points as we showed in Section III.A but which includes corrections due to these additional effects.

The full Hamiltonian describing all the effects discussed so far in the space of the eleven localized orbitals is given by
\begin{equation} H_{\mathbf{q}}^{\tau} = \left( {\begin{array}{cc}
H^{\textrm{E}} & H^{\textrm{EO}}_{} \\
H^{\textrm{EO}\dagger}_{} & H^{\textrm{O}} \\
\end{array}  } \right) \;,
\label{totalHamoddandeven}
\end{equation}
where the even sector includes mechanical deformations $H^{\textrm{E}} = H^{\textrm{E},\tau}_{\mathbf{q}} + \delta H^{\tau}_{}$ (given in Eq. (\ref{Fullham}) and Eq. (\ref{FullhamStrain}) respectively) acting in the space given by the first line of Eq. (\ref{basis}), the odd sector $H^{\textrm{O}}$ (given in Eq. (\ref{Fullham})) acts in the odd space shown in the second line of Eq. (\ref{basis}) while $H^{\textrm{EO}}_{}$ couples the even and odd sectors of the Hamiltonian and is given by $H^{\textrm{EO}}_{} = H_{\textrm{so},2} +  \delta H^{\tau}_{\textrm{curv}}$ (given in Eq. (\ref{SOCspinflip}) and Eq. (\ref{HamCurva}) respectively).

The procedure used to construct a low energy effective model contains two steps of successive Schrieffer-Wolff transformations.\cite{Schrieffer1966,Winker} First we fold down the full Hamiltonian in Eq. ({\ref{totalHamoddandeven}}) onto the even sub block to account for the effect of the even odd sector coupling mechanisms within the even sector of the Hamiltonian. Then we use another Schrieffer-Wolff transformation as outlined in Section II.A to produce the final low energy effective two-band Hamiltonian. We neglect the contributions which are second order in $\delta H^{\tau}_{\textrm{curv}}$, as these terms will only lead to corrections of order $(\nabla^2 h(\mathbf{r}))^2$ which for smooth ripples (where $h(\mathbf{r})$ changes slowly on the scale of the crystal lattice) will be small as compared to corrections arising due to bond length changes.

We calculate the effective six band Hamiltonian which describes the even sector with a Schrieffer-Wolff transformation as
\begin{align} H^{E,\tau}_{nm} \simeq &\; H^{\textrm{E},\tau}_{\mathbf{q},nm} + \delta H^{\tau}_{nm} \nonumber \\
&+ \frac{1}{2}\sum_{l}H^{\textrm{EO}}_{nl}H^{\textrm{EO}\dagger}_{lm}\Big[\frac{1}{\epsilon^{\textrm{E}}_{n}-\epsilon^{\textrm{O}}_{l}}+\frac{1}{\epsilon^{\textrm{E}}_{m}-\epsilon^{\textrm{O}}_{l}}\Big] \;, \end{align}
where $\epsilon^{\textrm{E}}_{n(m)}$ and $\epsilon^{\textrm{O}}_{l}$ refer to the eigenvalues of the even and odd sub blocks of the Hamiltonian respectively. Here we restrict the Hamiltonian in the odd sector $H^{\textrm{O}}_{\mathbf{q}}$ to the K and K$'$ points only, as corrections due to finite momenta near the K points in the higher energy odd sector lead only to small negligible corrects in the final effective Hamiltonian. 

We then proceed to find the effective Hamiltonian in the space of $(d^{\tau}_{z^2,\mathbf{q},s}, d^{\tau}_{\tau2,\mathbf{q},s})$ following the same procedure as outlined in Section II.A. This rather lengthy process finally yields   
\begin{align} H_{\mathbf{q},\textrm{eff}}^{E,\tau} = &\; H_{\textrm{0}} + H_{\textrm{trig}} + H_{\textrm{so}} \nonumber \\
& + \delta H + \delta H_{\textrm{so}} + \delta H_{\textrm{curv}} \;. \label{Hamlowstr} \end{align}
Here the first two terms $H_{\textrm{0}}$ and $H_{\textrm{trig}}$ are the same as in the low energy Hamiltonian unperturbed by the effects of strain and curvature presented in Eqs. (\ref{Ham2by20})-(\ref{Ham2by2trig}), whereas $H_{\textrm{so}}$ is modified by spin flip processes. Additionally we find $\delta H$ describing the coupling of mechanical deformations to orbital degrees of freedom, $\delta H_{\textrm{so}}$ contains terms describing the coupling of strain to spin degrees of freedom and $\delta H_{\textrm{curv}}$ describes the role of the coupling of curvature effects and spin degrees of freedom in the low energy theory. The expression presented in Eq. (\ref{Hamlowstr}) is the central result of this paper and we will now present each term in detail and discuss their role is the low energy physics of TMDCs.

Due to the inclusion of the spin flip terms which couple the even and odd sectors of the Hamiltonian the spin orbit Hamiltonian $H_{\textrm{so}}$ now contains new terms due to second order virtual transitions between the even and odd sector, additionally we will also present relevant sub leading term proportional to $\tilde{\lambda}_X$ which creates a spin splitting in the conduction band. The spin orbit interaction Hamiltonian is then given by 
\begin{equation} H_{\textrm{so}} = \; \left( {\begin{array}{cc}
\tau (\tilde{\lambda}_X - \lambda_{2})s_{z} & 0 \\
0 & 2 \tau \lambda_{M} s_{z} \\
\end{array} } \right) \;, \label{HamSoLow} \end{equation}
where $\lambda_{2}$ contains the corrections due to the second order spin flip transitions. 

The conduction band has predominately $d_{z^2,s}$ orbital character. Spin up electrons can transition to an odd band with $d_{\tau1,\downarrow}$ orbital character mediated by $\lambda_M$, whereas the spin down electrons may transition to an odd band consisting of a $d_{-\tau1,\uparrow}$ orbital state with the transition only being mediated by $\lambda_M$. Taking these processes into account and disregarding a negligible shift in the band splitting we find that $\lambda_{2}\simeq2\,\textrm{meV}$ in MoS$_2$.

To give a complete description of the spin splitting in the conduction band we must consider the direct splitting $\tilde{\lambda}_X$ arising from the p-orbitals contribution to the band and the corrections due to second order processes, these effects combined give a total spin splitting of $\simeq8\,\textrm{meV}$ in MoS$_2$. We must note that DFT calculations show that the conduction band does contain a minority contribution from the $p^S_{-\tau1,s}$ orbital\cite{Rossier2013,Burkard2014,Falko2015} and therefore for a more accurate prediction of the conduction band spin splitting a more detailed model is needed which correctly predicts the orbital weights of the bands. This feature of the band structure was not the focus of our study and has been studied in more detail by other authors.\cite{Rossier2013} Numerical studies of the conduction band spin splitting has predicted opposite splitting in the Tungsten based TMDCs, which will be achieved when the relation $\tilde{\lambda}_X < \lambda_{2}$ is satisfied.\cite{Burkard2014,Falko2015}

There also exists a correction to the spin splitting in the valence band. This occurs due to spin flip transitions between the valence band with an orbital structure of predominantly the $d_{\tau2,\downarrow}$ type orbital states and the odd symmetry band with an orbital makeup of $d_{-\tau1,\uparrow}$ states. The transitions are mediated by $\lambda_M$, but are second order corrections and as a consequence the strength of this process is several orders of magnitude smaller than the direct spin splitting, with the correction being $\simeq0.2\,\textrm{meV}$, therefore we neglect these corrections in our treatment. 
 
The terms describing the coupling of electronic and mechanical degrees of freedom can be expressed in two parts, and are most informatively expressed within $H_0$ as
\begin{equation} H_0 + \delta H = H_0 + \delta H_1 + \delta H_2 \;, \label{Hamdelta51} \end{equation}
where the first part contains the gauge fields which modifies the low energy Dirac-like terms in $H_0$. Here we present both as
\begin{align} H_0 + \delta H_1 = &\; \left( {\begin{array}{cc}
\Delta_c & vq + \eta_1 F^{(\tau)} \\
vq^{\dagger} + \eta_1 F^{(\tau)\dagger} & \Delta_v \\
\end{array}  } \right) \nonumber \\
& \; + \left( {\begin{array}{cc}
\beta\big| q+ \eta_2 F^{(\tau)}\big|^2 &  \kappa (q^{\dagger} + \eta_4 F^{(\tau)\dagger})^2 \\
 \kappa (q + \eta_4 F^{(\tau)})^2 & \alpha\big| q+ \eta_3 F^{(\tau)} \big|^2 \\
\end{array}  } \right) \label{Hamgauge}  \end{align}
with $F^{(\tau)} = \varepsilon_{yy} - \varepsilon_{xx} + 2i\tau \varepsilon_{xy}$. The coupling constant is found to be $\eta_1 = 150\textrm{meV}$, this weak coupling suggests that the role of the gauge fields is much weaker than in other similar systems such as graphene. In a very recent work this coupling constant was also obtained but was found to be two order of magnitude smaller than the $\eta_1$ presented here,\cite{Rostami2015} but nonetheless this shows agreement that this coupling is weak.  

The second line in Eq. (\ref{Hamgauge}) describes the gauge field corrections to the terms second order in momentum and will account for a tuning of the effective masses of the conduction and valence bands under deformations of the crystal lattice. The parameters are given by $\eta_2= 3.42\,\textrm{eV\AA$^{2}$}$, $\eta_3= 0.48\,\textrm{eV\AA$^{2}$}$ and $\eta_4=0.34\,\textrm{eV\AA$^{2}$}$. As a consequence strain offers a route to control the effective masses which play an important role defining the system in many situations such as electrical conductance, the cyclotron frequency of electron within a magnetic field and valley dependent g-factors which have been observed.\cite{Imamoglu2015,Xu2015,Ralph2015} 

The second section of Eq. (\ref{Hamdelta51}) gives the direct band edge shifts, given by
\begin{equation} \delta H_2 = \; \left( {\begin{array}{cc}
\delta_1 D & 0 \\
0 & \delta_2 D \\
\end{array}  } \right) + \left( {\begin{array}{cc}
\delta_3 D^2 & 0 \\
0 & \delta_4 D^2 \\
\end{array}  } \right) \;. \label{deltaH2}\end{equation} 


The term $\delta H_2$ includes the coupling of local lattice dilations to the diagonal electronic terms, where $D = \textrm{Tr}[\varepsilon_{ij}]$ is the local area variation. The terms in Eq. (\ref{deltaH2}) describe a tuning of the band edges under mechanical deformations, where the two linear term parameters are given by $\delta_1=-0.53\,\textrm{eV}$ and $\delta_2=-0.62\,\textrm{eV}$, and the terms which couple quadratically to the change in local area variation are $\delta_3=0.56\,\textrm{eV}$ and $\delta_4=0.02\,\textrm{eV}$. As a consequence of the broken inversion symmetry $\delta_1$ and $\delta_2$ are not required to be equal and therefore strain causes a decrease in the size of the band gap. Under an applied uniaxial strain of a form $\mathbf{u}(\mathbf{r}) = r \hat{r} \delta L/L$, with $\delta L / L$ describing the relative extension of the lattice, we find a decrease of the direct band gap at the K and K$'$ points $\simeq5\,\textrm{meV}/\%$. This value is approximately one order of magnitude smaller than values extracted from photoluminescence experiments,\cite{Bolotin2013} but it must be noted that the numerical values found in our calculations assume electron Gr\"{u}neisen parameters of unity. This direct band gap decrease has a strong dependence on variations in d-d bond lengths in $\gamma_1$ and $\gamma_2$ and also a weaker dependance on p-d bond lengths given in $\gamma_5$ and $\gamma_7$ suggesting electron Gr\"{u}neisen parameters larger than unity are necessary to explain current experiments. 
 
It is also interesting to note that there is an additional sub-leading term describing the coupling of spin to mechanical deformations of the lattice, which we give here for completeness. It is given by
\begin{equation} \delta H_{\textrm{so}} = s_z D\left( {\begin{array}{cc}
 \delta \lambda_1 & 0 \\
0 & \delta \lambda_2 \\
\end{array} } \right) \;, \end{equation}
where the local variations of area lead to an effective band dependent Zeeman field, and the couplings are given by $\delta \lambda_1 = 0.1\,\textrm{meV}$ and $\delta \lambda_2 = -0.02\,\textrm{meV}$. We see that dilations of the crystal lattice do indeed give rise to a shift in the size of the spin splitting in each band, but that the strength of the coupling is negligible. 

Now we turn our attention to the expression $\delta H_{\textrm{curv}}$ describing the curvature. The first consequence of curvature of the crystal lattice is that the induced curvature will locally produce a tilt as compared to the global coordinate system. In the local frame the electronic spin can be defined with respect to the local normal vector $\hat{\mathbf{z}}'(\mathbf{r})$ as $s_z = \mathbf{S}\cdot\hat{\mathbf{z}}'(\mathbf{r})$. Therefore now considering the global frame we find that this gives rise to a deflection coupling\cite{Rashba2010,Burkard2010}
\begin{equation}  s_z \rightarrow s_z - s_x \partial_x h - s_y \partial_y h  \;. \label{tilt} \end{equation}
This deflection coupling of the spin orbit interaction is an entirely geometrical effect and is due to a local tilt of the lattice that mixes out-of-plane deformations and in-plane electronic spins and will appear in Eq. (\ref{HamSoLow}). 

The second consequence of curvature of the crystal lattice is curvature changing the orbital compositions of the Bloch bands. Theses effects give rise to new terms in the effective low energy theory, given by 
\begin{equation} \delta H_{\textrm{curv}} =  \left( {\begin{array}{cc}
\tau\, \mathbf{S}\cdot\mathbf{B}_{\textrm{eff,c}} & \beta \nabla^2 h ( is_x+\tau s_y) \\
\beta \nabla^2 h (-is_x+\tau s_y) & \tau\, \mathbf{S}\cdot\mathbf{B}_{\textrm{eff,v}} \\
\end{array} } \right) \;. \label{HamCurv} \end{equation}
The diagonal terms describe the coupling of spin degrees of freedom to the local curvature of the crystal lattice.
Where the effect of the curvature is to create an effective in-plane magnetic field which couples with the spin degree of freedom of the electrons given by $\mathbf{S} = (s_x,s_y,s_z)$ a vector of the spin Pauli matrices. The effective magnetic fields which appear in the Hamiltonian are given by
\begin{align} \mathbf{B}_{\textrm{eff,c}} = &\; \big(\xi_1 2 \partial_x \partial_y h,\; \xi_2 (\partial^2_x h - \partial^2_y h),\; 0\big) \;, \nonumber \\
\mathbf{B}_{\textrm{eff,v}} = &\; \big(\xi_3 2 \partial_x \partial_y h,\; \xi_4 (\partial^2_x h - \partial^2_y h),\; 0\big) \;. \end{align}
Here we see that the role of the scalar height field $h(\mathbf{r})$ appears in the effective magnetic fields as components of the curvature tensor. This means that these expressions depend on second derivatives of the height field $h(\mathbf{r})$ and therefore are only sensitive to curvatures of the crystal lattice, in contrast to the coupling which appears in Eq. (\ref{tilt}). 

The parameters that govern the effective in-plane magnetic field in the conduction band $\mathbf{B}_{\textrm{eff,c}}$ are given by $\xi_1=115\,\textrm{meV\AA}$ and $\xi_2=67\,\textrm{meV\AA}$, whereas within the valence band effective in-plane magnetic field $\mathbf{B}_{\textrm{eff,v}}$ we find $\xi_3=19\,\textrm{meV\AA}$ and $\xi_4=12\,\textrm{meV\AA}$. We predict here that the effective in-plane magnetic field which couples to electronic spins in the conduction band is two orders of magnitude larger than the analogous effect in the valence band.
 
The magnitude of the effective magnetic field is equal and opposite in each of the valleys, therefore time reversal symmetry is broken locally within each valley and not globally over the entire Brillouin zone. This is expected as elastic deformations of the lattice are time reversal invariant. As a consequence of this we would expect spin transport scattering due to ripples to depend strongly on whether there is strong intra-valley scattering present which breaks time reversal symmetry or strong inter-valley scattering processes which are time reversal invariant. 

The off diagonal terms within $\delta H_{\textrm{curv}}$ represent inter-band spin-lattice coupling mechanisms. The strength of the coupling is given by $\beta=105\,\textrm{meV\AA}$. This term shows the direct coupling of electronic spins to the out-of-plane deformations given by the mean curvature of the surface that describes the crystal lattice. This coupling is analogous to a Rashba type spin orbit interaction, being of the form $\propto \tau \sigma_x s_y - \sigma_y s_x$, where here $\sigma_j$ are the Pauli matrices acting in the band space.

There also exists off diagonal terms which couple an effective in-plane magnetic field $\mathbf{B}_{\textrm{eff,vc}}$ with the spin degrees of freedom. These terms only provide sub-leading corrections to the Hamiltonian with coupling parameters of order $\simeq5\,\textrm{meV\AA}$, much smaller than other previously mentioned off diagonal terms and therefore we neglect these terms within our treatment. 

An experimental study preformed by Brivio {\it et al}\cite{Kis2011b} looked at the structure of the spontaneous ripples formed in MoS$_2$. With the use of high-resolution transmission electron microscopy and atomic force microscopy they observe that typical ripples have lengths of $6-10\,\textrm{nm}$ and heights of $6-10\,\textrm{\AA}$, and these observations have been supported by molecular dynamics simulations.\cite{Heine2013b} Therefore, based on dimensional analysis, we would expect the magnitude of components of the curvature tensor for typical ripples to be of order $\simeq10^{-3}\,\textrm{\AA}^{-1}$. Comparing this to the Rashba spin-orbit coupling which arises due to perpendicular electric fields, we see that curvature effects can have a comparable effect for realistic magnitudes of electric fields.\cite{Burkard2014} 

\section{Conclusions}

In this work we have presented a tight binding study of the TMDCs and have included the effects of both mechanical deformations which cause bond length changes and curvature which leads to a mixing of the orbital structure of the Bloch bands. We find that mechanical deformations allow for tuning of the direct band gap at the K  and K$'$ points and lead to fictitious gauge fields which couple to the orbital degrees of freedom. In parallel curvature breaks the mirror inversion symmetry of the crystal lattice and introduces spin-lattice coupling mechanisms that manifest themselves as an effective in-plane magnetic field which couples to spin and a Rashba like coupling with a magnitude proportional to the mean curvature of the TMDC lattice. Due to the microscopic nature of the model presented we have produced estimates of the strengths of the coupling parameters of all the relevant processes.

This study provides a basis for further study into a microscopic understanding of the role of strain on TMDC monolayers and of future devices which utilize the coupling of mechanical and electronic degrees of freedom allowing for the tailoring of electronic band and spin-orbit coupling strengths to suit device needs.  

\acknowledgements

We thank V.\ Z\'{o}lyomi for sharing density functional theory data with us and A.\ Korm\'{a}nyos for fruitful discussions. AJP and GB acknowledge funding from DFG under the program SFB 767 and the EU through Marie Curie ITN S$^3$NANO. EM acknowledges the financial support of the Royal Society (International Exchange Grant No. IE140367) and of the Leverhulme Trust (Research Project Grant RPG-2015-101).

\appendix

\section{Tight Binding parameters}

\subsection{Parameters of the Six Band Tight Binding Model}

In this section we present the form of the terms in the six band tight binding model presented in Section III, Eq's (\ref{HamKpointdd})-(\ref{HamKpointpd}). The energies and group velocities which appear in the Hamiltonian are given in terms of the on-site energies and Koster-Slater parameters discussed in the derivation of the model in Section II. These expressions are written as
\begin{align} V_{D0}(\mathbf{P}) = &\; \epsilon_{d}^0 - \frac{3}{4}(3V_{dd}^{\delta}+V_{dd}^{\sigma})\mathbf{P} \\
 V^{(\pm)}_{D2}(\mathbf{P}) = &\; V_{D2}(\mathbf{P}) \pm 2\lambda_{M}\tau s_{z}\\
 V_{D2}(\mathbf{P}) = &\; \epsilon_{d}^2 - \frac{3}{8}(V_{dd}^{\delta}+4V_{dd}^{\pi}+3V_{dd}^{\sigma})\mathbf{P} \\
 V^{(\pm)}_{P1}(\mathbf{P}) = &\; V_{P1}(\mathbf{P}) \pm \lambda_{X} \tau s_{z} \\
 V_{P1}(\mathbf{P}) = &\; \epsilon^1_p + V_{pp}^{\pi} - \frac{3}{2}(V_{pp}^{\pi}+V_{pp}^{\sigma})\mathbf{P} \\
 V_{P0}(\mathbf{P}) = &\; \epsilon^0_p - V_{pp}^{\sigma} - 3 V_{pp}^{\pi}\mathbf{P} \\
 K_{pd}^{(1)} = &\; \frac{3a}{2\tilde{c}^3}(2c^2(\sqrt{3}V_{pd}^{\pi}-V_{pd}^{\sigma})+a^2V_{pd}^{\sigma}) \\
 K_{pd}^{(2)} = &\; \frac{3a^2c}{2\tilde{c}^3}(2V_{pd}^{\pi}-\sqrt{3}V_{pd}^{\sigma}) \\
 K_{pd}^{(2s)} = &\; \frac{3}{2\sqrt{2}\tilde{c}^3}(a^3(\sqrt{3}V_{pd}^{\sigma}-2V_{pd}^{\pi})+4a\tilde{c}^2V_{pd}^{\pi}) \\
 v_{dd}^{(0)} = &\; \frac{9\sqrt{3}a}{8\sqrt{2}}(V_{dd}^{\delta}-V_{dd}^{\sigma}) \\
 v_{dd}^{(2)} = &\; \frac{9a}{16} (V_{dd}^{\delta}-4V_{dd}^{\pi}+3V_{dd}^{\sigma}) \\
 v_{pd}^{(0)} = &\; \frac{3ac}{2\sqrt{2}\tilde{c}^3}(a^2(2\sqrt{3}V_{pd}^{\pi}-V_{pd}^{\sigma})+2c^2V_{pd}^{\sigma}) \\
 v_{pd}^{(1)} = &\; \frac{aK_{pd}^{(1)}}{2} \\
 v_{pd}^{(2)} = &\; \frac{aK_{pd}^{(2)}}{2} \\
 v_{pd}^{(2s)} = &\; \frac{aK_{pd}^{(2s)}}{2} \\
 v_{pd}^{(2a)} = &\; \frac{3}{4\sqrt{2}\tilde{c}^3}(a^4(2V_{pd}^{\pi}-\sqrt{3}V_{pd}^{\sigma}) \\
 v_{pp} = &\; \frac{9a}{4}(V_{pp}^{\pi}-V_{pp}^{\sigma}) \;. \end{align}
 
Where $\mathbf{P} = 1 - (3a^2/4)|\mathbf{q}|^2$. Comparing these expressions with first principle calculations we present the numerical values of the parameters of the six band model for the case of $\mathbf{q}=0$ in Table I.

\begin{center}
\begin{table}
\caption{The numerical values of the parameters for MoS$_2$ which appear in the six band Hamiltonian. The parameters are given for the case at the band edges, in which $\mathbf{q}=0$. All energies are given in units of eV and group velocities in units of eV$\textrm{\AA}$}
\begin{tabular}{ c  c | c  c | c  c} \hline 
\hline \hline
 $V_{D0}$ & 1.28 & $K_{pd}^{(1)}$ & 1.79 & $v_{pd}^{(0)}$ & 1.01 \\ 
 $V_{D2}$ & -0.37 & $K_{pd}^{(2)}$ & 3.59 & $v_{pd}^{(1)}$ & 1.64 \\
 $V_{P1}$ & -4.09 & $K_{pd}^{(2s)}$ & -0.89 & $v_{pd}^{(2)}$ & 3.30 \\ 
 $V_{P0}$ & 0.42 & $v_{dd}^{(0)}$ & 1.77 & $v_{pd}^{(2s)}$ & -0.81 \\ 
$v_{pp}$ & -3.75 & $v_{dd}^{(2)}$ & 2.28 & $v_{pd}^{(2a)}$ & 2.85 \\ 
\hline \hline
\end{tabular}
\end{table}
\end{center}
\subsection{Fitting to First Principle Calculations}

We find the on-site energies and Koster-Slater parameters of the six band Hamiltonian by fitting to DFT calculations of the electronic band structure of MoS$_2$. This is done by minimizing the function 
\begin{equation} f(\mathbf{q}) = \sum_{n,\mathbf{q}} \big( \epsilon^{\textrm{TB}}_n(\mathbf{q}) - \epsilon_n^{\textrm{DFT}}(\mathbf{q}) \big)^2 \label{mini}\end{equation}  
where $\epsilon^{\textrm{TB}}_n(\mathbf{q})$ corresponds to the eigenvalues of the n-th band of the tight binding Hamiltonian and $\epsilon^{\textrm{DFT}}_n(\mathbf{q})$ is the energy of the n-th band found via DFT techniques. We minimize Eq (\ref{mini}) using the Powell method fitting procedure.\cite{Powell64} Firstly a fitting is preformed at the band edges which are used to fix the onsite energies, then we fit a range of $\approx 0.1 (2\pi/a)$ momenta around the K point in the $\Gamma$ to K to M direction to find the Koster-Slater hopping parameters. To achieve a good fitting it is necessary to extend the expansion of $\mathbf{q}$ around the K points to second order. We fit to band structures calculated without the effect of spin-orbit interactions included, and leave a more exhaustive fitting including spin-orbit interactions to a later study. 

The Koster-Slater parameters resulting from this procedure can be found in Table II. Where we present the numerical values for the energies and group velocities of our six band model found during this fitting procedure. 

\begin{center}
\begin{table}
\caption{The numerical values of the onsite energies and Koster-Slater parameters which appear in the six band Hamiltonian for MoS$_2$. All energies are given in units of eV}
\begin{tabular}{ l | c | r } 
\hline \hline
On-site Energies & $\epsilon^{0}_p$ & -3.96 \\ 
 & $\epsilon^1_p$ & -5.38 \\ 
 & $\epsilon_d^0$ & 2.12 \\
 & $\epsilon_d^1$ & -0.46 \\
 & $\epsilon_d^2$ & -1.41 \\ 
S-S hoppings & $V_{pp}^{\pi}$ & -1.32 \\ 
 & $V_{pp}^{\sigma}$ & -0.42 \\ 
S-Mo hoppings & $V_{pd}^{\pi}$ & 0.67 \\ 
 & $V_{pd}^{\sigma}$ & -2.83 \\
Mo-Mo hoppings & $V_{dd}^{\delta}$ & 0.45 \\ 
 & $V_{dd}^{\pi}$ & -0.62 \\ 
 & $V_{dd}^{\sigma}$ & -0.24 \\ 
\hline \hline
\end{tabular}
\end{table}
\end{center}

\subsection{Odd Sector of the Tight Binding Hamiltonian around the K point}

In this section of the appendix we present the full form of the odd sector of the Hamiltonian near the K point presented in Eq. (\ref{Fullham}) in the basis of the odd states shown in Eq. (\ref{basis}). We expand to linear order in $\mathbf{q}$, where $\mathbf{q}$ is the momentum close to the K Point given by $\mathbf{k} = \mathbf{k}_D + \mathbf{q}$. This process yields
\begin{align} H_{\mathbf{q},dd}^{\textrm{O},\tau} = &\, \left( {\begin{array}{ccc}
V^{(+)}_{D1} & -v_{dd}^{(1)}q^{\dagger} \\
-v_{dd}^{(1)}q & V^{(-)}_{D1} \\
\end{array}  } \right) 
\label{HamGpointdd} \;, \\
H_{\mathbf{q},pp}^{\textrm{O},\tau} = &\, \left( {\begin{array}{ccc}
V^{(+)\textrm{O}}_{P1} & -v_{pp}q & 0 \\
-v_{pp}q^{\dagger} & V^{(-)\textrm{O}}_{P1} & 0 \\
0 & 0 & V_{P0}^{\textrm{O}} \\
\end{array}  } \right) \;,
\label{HamGpointpp} \\
H_{\mathbf{q},pd}^{\textrm{O},\tau} = &\, \left( {\begin{array}{ccc} 
v_{pd}^{(1s)}q^{\dagger} & \sqrt{2}v_{pd}^{(2)}q & i\tau K_{pd}^{(1z)} \\
\sqrt{2} K_{pd}^{(2)} & - v_{pd}^{(1s)}q^{\dagger} & -i\tau v_{pd}^{(1z)}q \\
\end{array}  } \right) \,.
\label{HamGpointpd} \end{align}
Where we have introduced the new on-site energies and group velocities 
\begin{align} V_{D1}^{(\pm)} = &\; V_{D1} \pm \lambda_{M} \tau s_{z} \\
V_{D1} = &\; \epsilon_{d}^1 - \frac{3}{2}(V_{dd}^{\delta}+V_{dd}^{\pi}) \\
V_{P1}^{(\pm)\textrm{O}} = &\; V_{P1}^{\textrm{O}} \pm \lambda_{X} \tau s_{z} \\
V_{P1}^{\textrm{O}} = &\; \epsilon^1_p - V_{pp}^{\pi} - \frac{3}{2}(V_{pp}^{\pi}+V_{pp}^{\sigma})  \\
V_{P0}^{\textrm{O}} = &\; \epsilon^0_p + V_{pp}^{\sigma} -3V_{pp}^{\pi} \\
K_{pd}^{(1z)} = &\; \frac{3a}{\tilde{c}^3}(c^2(\sqrt{3}V_{pd}^{\sigma}-2V_{pd}^{\pi})+\tilde{c}^2V_{pd}^{\pi}) \\
v_{dd}^{(1)} = &\; \frac{9a}{4}(V_{dd}^{\delta}-V_{dd}^{\pi}) \\
v_{pd}^{(1s)} = &\; \frac{3c}{2\sqrt{2}\tilde{c}^3} (a^3(\sqrt{3}V_{pd}^{\sigma}-2V_{pd}^{\pi})+2a\tilde{c}^2V_{pd}^{\pi}) \\
v_{pd}^{(1z)} = &\; \frac{K_{pd}^{(1z)}a}{2} \end{align}
Here we see that at precisely the band edge ($\mathbf{q}=0$) each d-orbital couples to one p-orbital while the $p^{A}_{\mathbf{k},-\tau1,s}$ remains uncoupled, this is in excellent agreement with other tight binding models.\cite{Guinea2013}

The numerical values of the parameters of the odd sector Hamiltonian are presented in Table III.
\begin{center}
\begin{table}
\caption{The numerical values of the parameters for MoS$_2$ which appear in the five band Hamiltonian of the odd sector of the full tight binding Hamiltonian. All energies are given in units of eV and group velocities in units of eV$\textrm{\AA}$}
\begin{tabular}{ c  c | c  c | c  c} \hline 
\hline \hline
 $V_{D1}$ & -0.21 & $K_{pd}^{(1z)}$ & -4.33 & $v_{dd}^{(1)}$ & 4.45 \\ 
 $V_{P1}^{\textrm{O}}$ & -1.45 & & & $v_{pd}^{(1s)}$ & -3.01 \\
 $V_{P0}^{\textrm{O}}$ & -0.41 & & & $v_{pd}^{(1z)}$ & -3.99 \\ 
 \hline \hline
\end{tabular}
\end{table}
\end{center}

\section{Parameters Used in the Analysis of Strain}

\subsection{Parameters for Mechanical Deformations}

In this section we present the couplings $\gamma_k$ which arise in the strain induced corrections to the six band Hamiltonian shown in Eq. (\ref{FullhamStrain}) which are given by
\begin{align} \gamma_1 = & \; \frac{3}{8}\big[3\Gamma_{dd}^{\delta}V_{dd}^{\delta} + \Gamma_{dd}^{\sigma}V_{dd}^{\sigma}\big] \\
\gamma_2 = & \; \frac{3}{16}\big[\Gamma_{dd}^{\delta}V_{dd}^{\delta} + 4\Gamma_{dd}^{\pi}V_{dd}^{\pi} + 3\Gamma_{dd}^{\sigma}V_{dd}^{\sigma}\big] \\
\gamma_3 = & \; -\frac{3}{32}\big[\Gamma_{dd}^{\delta}V_{dd}^{\delta} - 4\Gamma_{dd}^{\pi}V_{dd}^{\pi} + 3\Gamma_{dd}^{\sigma}V_{dd}^{\sigma}\big] \\
\gamma_4 = & \; -\frac{3}{16}\sqrt{\frac{3}{2}}\big[\Gamma_{dd}^{\delta}V_{dd}^{\delta} - \Gamma_{dd}^{\sigma}V_{dd}^{\sigma}\big] \\
\gamma_5 = & \; -\frac{3a^3}{4\tilde{c}^5}\Big[c^2\big[\sqrt{3}\Gamma_{pd}^{\pi}V_{pd}^{\pi} - \Gamma_{pd}^{\sigma}V_{pd}^{\sigma}\big] - \frac{a^2}{2}\Gamma_{pd}^{\sigma}V_{pd}^{\sigma} \Big]\\
\gamma_6 = & \; \frac{3a^4c}{4\sqrt{2}\tilde{c}^5}\Big[\big[2\sqrt{3}\Gamma_{pd}^{\pi}V_{pd}^{\pi} - \Gamma_{pd}^{\sigma}V_{pd}^{\sigma}\big] + \frac{\sqrt{2}c^2}{a\tilde{c}} \Gamma_{pd}^{\sigma}V_{pd}^{\sigma}\Big] \\
\gamma_7 = & \; \frac{3a^5}{4\sqrt{2}\tilde{c}^5}\Big[\Gamma_{pd}^{\pi}V_{pd}^{\pi} - \frac{\sqrt{3}}{2} \Gamma_{pd}^{\sigma}V_{pd}^{\sigma} - \frac{2\tilde{c}^2}{a^2}\Gamma_{pd}^{\pi}V_{pd}^{\pi}\Big] \\
\gamma_8 = & \; -\frac{3a^4c}{8\tilde{c}^5}\Big[2 \Gamma_{pd}^{\pi}V_{pd}^{\pi} - \sqrt{3} \Gamma_{pd}^{\sigma}V_{pd}^{\sigma}\Big] \\
\gamma_9 = & \; -\frac{a}{c\sqrt{2}}\gamma_8 \\
\gamma_{10} = & \; -\frac{3}{4} \big[ \Gamma_{pp}^{\sigma}V_{pp}^{\sigma} + \Gamma_{pp}^{\pi}V_{pp}^{\pi} \big] \\
\gamma_{11} = & \; \frac{3}{4} \Gamma_{pp}^{\pi} V_{pp}^{\pi} \\
\gamma_{12} = & \; \frac{3}{8} \big[ \Gamma_{pp}^{\sigma}V_{pp}^{\sigma} - \Gamma_{pp}^{\pi}V_{pp}^{\pi} \big] \,.
\end{align}
The numerical values of the parameters $\gamma_k$  are shown in Table IV. For the example case of MoS$_2$, the values presented are calculated under the assumption that the electron Gr\"{u}neisen parameters $\Gamma_{dd}^{\mu}$, $\Gamma_{pd}^{\mu}$ and $\Gamma_{pp}^{\mu}$ are taken to be unity.

\begin{center}
\begin{table}
\caption{The numerical values of the coupling parameters which appear in the strain induced corrections to the six band Hamiltonian for the case of MoS$_2$. The electron Gr\"{u}neisen parameters are all taken to be one. All energies are given in units of eV.}
\begin{tabular}{ c  c | c  c | c c | c c} \hline 
\hline \hline
 $\gamma_1$ & 0.42 & $\gamma_4$ & -0.16 & $\gamma_7$ & 0.13 & $\gamma_{10}$ & 1.31 \\
 $\gamma_2$ & -0.52 & $\gamma_5$ & -0.86 & $\gamma_8$ & -0.54 & $\gamma_{11}$ & -0.99 \\
 $\gamma_3$ & -0.21 & $\gamma_6$ & -1.46 & $\gamma_9$ & 0.47 & $\gamma_{12}$ & 0.34 \\
\hline \hline
\end{tabular}
\end{table}
\end{center}

\subsection{Parameters for Curvature}

In this section we present the relevant parameters $V_j$ which arise in the description of the curvature which couples the odd and even sections of the unperturbed Hamiltonian. These terms appear in the Hamiltonian shown in Eq. (\ref{HamCurva}) and are given by
\begin{align} V_1 =&\; \frac{9}{8}\sqrt{\frac{3}{2}}\big[ V_{dd}^{\delta} - 2V_{dd}^{\pi} + V_{dd}^{\sigma}\big] \\
V_2 =&\; \frac{27}{16}\big[V_{dd}^{\delta} - V_{dd}^{\sigma}\big] \\
V_3 =&\; \frac{9}{8}\big[V_{dd}^{\delta} - 4 V_{dd}^{\pi} + 3 V_{dd}^{\sigma}\big] \\
V_4 =&\; \frac{3c}{4\tilde{c}^3}\big[a^2(4\sqrt{3}V_{pd}^{\pi} - 7V_{pd}^{\sigma})+2c^2(V_{pd}^{\sigma}-\sqrt{3}V_{pd}^{\pi})\big] \\
V_5 =&\; \frac{3a}{4\tilde{c}^3}\big[a^2(2\sqrt{3}V_{pd}^{\pi} - V_{pd}^{\sigma})-4c^2(\sqrt{3}V_{pd}^{\pi}-2V_{pd}^{\sigma})\big]\\
V_6 =&\; \frac{3c}{4\sqrt{2}\tilde{c}^3}\big[a^2(6V_{pd}^{\pi}-3\sqrt{3}V_{pd}^{\sigma})-4\tilde{c}^2V_{pd}^{\pi}\big]\\
V_7 =&\; \frac{9a^2c}{4\sqrt{2}\tilde{c}^3}\big[2V_{pd}^{\pi}-\sqrt{3}V_{pd}^{\sigma}\big]\\
V_8 =&\; \frac{3a}{4\tilde{c}^3}(a^2-2c^2)\big[2V_{pd}^{\pi}-\sqrt{3}V_{pd}^{\sigma}\big]\\
V_9 =&\; \sqrt{2}V_8 - \frac{3a}{2\sqrt{2}\tilde{c}}V_{pd}^{\pi} \\
V_{10} =&\; 2V_8 + \frac{3c}{2\tilde{c}}V_{pd}^{\pi} \;.\end{align} 
Once again we present the numerical values of the parameters $V_j$ in Table V.

\begin{center}
\begin{table}[!ht]
\caption{The numerical values of the coupling parameters which appear in the Hamiltonian describing curvature shown in Eq. (\ref{HamCurva}) for the material parameters of MoS$_2$. All energies are given in units of eV.}
\begin{tabular}{ c  c | c  c | c c | c c | c c} \hline 
\hline \hline
 $V_1$ & 1.17 & $V_2$ & 0.83 & $V_3$ & 1.36 & $V_4$ & 5.43 & $V_5$ & -3.35 \\
 $V_6$ & 2.27 & $V_7$ & 3.99 & $V_8$ & -0.79 & $V_9$ & -2.17 & $V_{10}$ & -0.37 \\
\hline \hline
\end{tabular}
\end{table}
\end{center}


\begin{thebibliography}{99}

\bibitem{Strano2012} Q.H.\ Wang, K.\ Kalantar-Zadeh, A.\ Kis, J.N.\ Coleman and M.S.\ Strano, Nat. Nanotech. {\bf7}, 699 (2012).

\bibitem{Heinz2010} K.F.\ Mak, C.\ Lee, J.\ Hone, J.\ Shan and T.F.\ Heinz, Phys. Rev. Lett.  {\bf105}, 136805 (2010).

\bibitem{Xiao2012} D.\ Xiao, G-B.\ Liu, W.\ Feng, X.\ Xu and W.\ Yao, Phys. Rev. Lett. {\bf108}, 196802 (2012).

\bibitem{Heinz2012} K.F.\ Mak, K.\ He, J.\ Shan and T.F.\ Heinz, Nat. Nanotech. {\bf7}, 494 (2012).

\bibitem{Kis2011} B.\ Radisavljevic, A.\ Radenovic, J.\ Brivio, V.\ Giacometti and A.\ Kis, Nat. Nanotech. {\bf6}, 147-150 (2011).

\bibitem{Falko2015} A.\ Korm\'{a}nyos, G.\ Burkard, M.\ Gmitra, J.\ Fabian, V.\ Z\'{o}lyomi, N.D.\ Drummond and V.\ Fal'ko, 2D Mater. {\bf2}, 022001 (2015).

\bibitem{Bollinger2012} A.\ Castellanos-Gomez, M.\ Poot, G.A.\ Steele, H.S.J.\ van der Zant, N.\ Agra\:{i}t and G.\ Rubio-Bollinger, Adv. Mater. {\bf24}, 772-775 (2012).

\bibitem{Venstra2013} A.\ Castellanos-Gomez, R.\ van Leeuwen, M.\ Buscema, H.S.J.\ van der Zant, G.A.\ Steele and W.J.\ Venstra Adv. Mater. {\bf25}, 6719-6723 (2013).

\bibitem{Lee2013} T.\ Jin, J.\ Kang, E.S.\ Kim, S.\ Lee and C.\ Lee, J. Appl. Phys. {\bf114}, 164509 (2013).

\bibitem{Huang2013} H.\ Shi, R.\ Yan, S.\ Bertolazzi, J.\ Brivio, B.\ Gao, A.\ Kis, D.\ Jena, H.G.\ Xing and L.\ Huang, ACS Nano. {\bf7}, 1072-1080 (2013).

\bibitem{Ando2002} H.\ Suzuura and T.\ Ando, Phys. Rev. B {\bf65}, 235412 (2002).

\bibitem{Guinea2010} M.A.H.\ Vozmediano, M.I.\ Katsnelson and F.\ Guinea, Physics Reports {\bf496}, 109-148 (2010).

\bibitem{Ando2000} T.\ Ando, J. Phys. Soc. Jpn {\bf69}, 1757-1763 (2000).  

\bibitem{Brataas2006} D.\ Huertas-Hernando, F.\ Guinea and A.\ Brataas, Rhys. Rev. B, {\bf74}, 155426 (2006). 

\bibitem{Lee2011} J-S.\ Jeong, J.\ Shin and H-W.\ Lee, Phys. Rev. B, {\bf84}, 195457 (2011).

\bibitem{Ochoa2013} H.\ Ochoa, F.\ Guinea, and V.I.\ Fal'ko, Phys. Rev. B {\bf88}, 195417 (2013).

\bibitem{Kis2011b} J.\ Brivio, D.T.L.\ Alexander and A.\ Kis, Nano Lett. {\bf11}, 5148-5153 (2011).

\bibitem{Geim2008} M.I.\ Katsnelson and A.K.\ Geim, Phil. Trans. R. Soc. A  {\bf366}, 195-204 (2008).

\bibitem{Roldan2013} H.\ Ochoa and R.\ Rold\'{a}n, Phys. Rev. B {\bf87}, 245421 (2013).

\bibitem{Dery2013} Y.\ Song and H.\ Dery, Phys. Rev. Lett. {\bf111}, 026601 (2013).

\bibitem{Ochoa2014} H.\ Ochoa, F.\ Finocchiaro, F.\ Guinea and V.I.\ Fal'ko, Phys. Rev. B, {\bf90}, 235429 (2014).

\bibitem{Zhang2012} H.\ Pan, Y-W.\ Zhang, The Journal of Physical Chemistry C, {\bf116}, 11752-11757, (2012).

\bibitem{Zeng2012} P.\ Lu, X.\ Wu, W.\ Guo, X.C.\ Zeng, Physical chemistry chemical physics PCCP, {\bf14}, 13035-40, (2012).

\bibitem{Li2012} Q.\ Yue, J.\ Kang, Z.\ Shao, X.\ Zhang, S.\ Chang, G.\ Wang, S.\ Qin, J.\ Li, J. Physics Letters A, {\bf376}, 1166-1170, (2012).

\bibitem{Shenoy2012} P.\ Johari and V.B.\ Shenoy, ACS Nano. {\bf6}, 5449 (2012).

\bibitem{Walle2012} H.\ Peelaers, C.G.\ Van de Walle, Phys. Rev. B, {\bf86}, 241401(R) (2012). 

\bibitem{Heine2013} M.\ Ghorbani-Asl, S.\ Borini, A.\ Kuc, T.\ Heine, Phys. Rev. B, {\bf87}, 235434 (2013).

\bibitem{Steele2013} A.\ Castellanos-Gomez, R.\ Rold\'{a}n, E.\ Cappelluti, M.\ Buscema, F.\ Guinea, H.S.J. van der Zant and G.A.\ Steele, Nano. Lett. {\bf13}, 5361 (2013).
 
\bibitem{Shan2013} K.\ He, C.\ Poole, K.F.\ Mak, and J.\ Shan, Nano. Lett. {\bf13}, 2931 (2013). 

\bibitem{Bolotin2013} H.J.\ Conley, B.\ Wang, J.I.\ Ziegler, R.F.\ Haglund, Jr., S.T.\ Pantelides, and K.I.\ Bolotin, Nano. Lett. {\bf13}, 3626 (2013). 

\bibitem{Urbaszek2013} C.R.\ Zhu, G.\ Wang, B.L.\ Liu, X.\ Marie, X.F.\ Qiao, X. Zhang, X.X.\ Wu, H.\ Fan, P.H.\ Tan, T.\ Amand and B.\ Urbaszek, Phys. Rev. B {\bf88}, 121301(R) (2013).

\bibitem{Javey2014} S.B.\ Desai, G.\ Seol, J.S.\ Kang, H.\ Fang, C.\ Battaglia, R.\ Kapadia, J.W.\  Ager, J.\ Guo and A.\ Javey, Nano. Lett. {\bf14}, 4592 (2014).

\bibitem{Wang2014} W.\ Wu, L.\ Wang, Y.\ Li, F.\ Zhang, L.\ Lin, S.\ Niu, D.\ Chenet, X.\ Zhang, Y.\ Hao, T.F.\ Heinz, J.\ Hone and Z.L.\ Wang, Nature {\bf514}, 470-474 (2014).

\bibitem{Zhang2015} H.\ Zhu, Y.\ Wang, J.\ Xiao, M.\ Liu, S.\ Xiong, Z.J.\ Wong, Z.\ Ye, Y.\ Ye, X.\ Yin and X.\ Zhang, Nat. Nanotech. {\bf10}, 151-155 (2015).

\bibitem{Koster1954} J.C.\ Slater and G.F.\ Koster, Phys. Rev. {\bf94}, 1498 (1954).

\bibitem{Harrison} W.A.\ Harrison, {\it Elementary Electronic Structure} (World Scientific Publishing, 1999).

\bibitem{Eriksson2009} S.\ Leb\'{e}gue and O.\ Eriksson, Phys. Rev. B, {\bf79}, 115409 (2009).

\bibitem{Schwingenschlogl2011} Z.Y.\ Zhu, Y.C.\ Cheng and U.\ Schwingenschl\"{o}gl, Phys. Rev. B {\bf84}, 153402 (2011).
 
\bibitem{Falko2013} A.\ Korm\'{a}nyos, V.\ Z\'{o}lyomi, N.D.\ Drummond, P.\ Rakyta, G.\ Burkard and V.I.\ Fal'ko, Phys. Rev. B, {\bf88}, 045416 (2013).

\bibitem{Guinea2013} E.\ Cappelluti, R.\ Rold\'{a}n, J.A.\ Silva-Guill\'{e}n, P.\ Ordej\'{o}n and F.\ Guinea, Phys. Rev. B, {\bf88}, 075409 (2013).

\bibitem{Imamoglu2015} A.\ Srivastava, M.\ Sidler, A.V.\ Allain, D.S.\ Lembke, A.\ Kis and A.\ Imamo\v{g}lu, Nat. Phys. {\bf11}, 141-147 (2015).

\bibitem{Xu2015} G.\ Aivazian, Z.\ Gong, A.M.\ Jones, R-L.\ Chu, J.\ Yan, D.G.\ Mandrus, C.\ Zhang, D.\ Cobden, W.\ Yao and X.\ Xu, Nat. Phys. {\bf11} 148-152 (2015).

\bibitem{Ralph2015} D.\ MacNeill, C.\ Heikes, K.F.\ Mak, Z.\ Anderson, A.\ Korm\'{a}nyos, V.\ Z\'{o}lyomi, J.\ Park, and D.C.\ Ralph, Phys. Rev. Lett. {\bf114}, 037401 (2015).

\bibitem{Rossier2013} K. Ko\'{s}mider, J.W.\ Gonz\'{a}lez and J.\ Fern\'{a}ndez-Rossier, Phys. Rev. B, {\bf88}, 245436 (2013).

\bibitem{Fang2015} S.\ Fang, R.\ Kuate Defo, S.N.\ Shirodkar, S.\ Lieu, G.A.\ Tritsaris, E.\ Kaxiras, Phys. Rev. B, {\bf92}, 205108 (2015).

\bibitem{Ridolfi2015} E.\ Ridolfi, D.\ Le, T.S.\ Rahman, E.R.\ Mucciolo, and C.H.\ Lewenkopf, J. Phys. Condens. Matter. {\bf27}, 365501 (2015).

\bibitem{Schrieffer1966} R.\ Schrieffer and P.A.\ Wolff, Phys. Rev. {\bf149}, 491 (1966).

\bibitem{Winker} R. Winkler, {\it Spin-Orbit Coupling Effects in Two-Dimensional Electron and Hole Systems} (Springer, Berlin, 2003).

\bibitem{Liu2013} G-B.\ Liu, W-Y.\ Shan, Y.\ Yao, W.\ Yao, and D.\ Xiao, Phys. Rev. B {\bf88}, 085433 (2013).

\bibitem{Asgari2013} H.\ Rostami, A.G.\ Moghaddam and R.\ Asgari, Phys. Rev. B {\bf88}, 085440 (2013).

\bibitem{Burkard2014} A.\ Korm\'{a}nyos, V.\ Z\'{o}lyomi, N.D.\ Drummond and G.\ Burkard, Phys. Rev. X, {\bf4}, 011034 (2014).

\bibitem{Landaubook} L.D.\ Landau and E.M.\ Lifshitz, {\it Theory of Elasticity} (Pergamon, New York, 1986).

\bibitem{Oppen2012} E.\ Mariani, A.J.\ Pearce and F.\ von Oppen, Phys. Rev. B {\bf86}, 165448 (2012).

\bibitem{Guinea2014} M.A.\ Cazalilla, H.\ Ochoa and F.\ Guinea, Phys. Rev. Lett, {\bf113}, 077201 (2014).

\bibitem{Crommie2010} N.\ Levy, S.A.\ Burke, K.L.\ Meaker, M.\ Panlasigui, A.\ Zettl, F.\ Guinea, A.H.\ Castro Neto and M.F.\ Crommie, Science, {\bf329}, 544 (2010).

\bibitem{diffgeobook} D.J.\ Struik {\it Lectures on Classical Differential Geometry}, 2nd Ed. (Dover, New York, 1988).

\bibitem{Rostami2015} H.\ Rostami, R.\ Rold\'{a}n, E.\ Cappelluti, R.\ Asgari, and F.\ Guinea, Phys. Rev. B {\bf 92}, 195402 (2015).

\bibitem{Rashba2010} M.S.\ Rudner and E.I.\ Rashba, Phys. Rev. B {\bf81}, 125426 (2010).

\bibitem{Burkard2010} P.R.\ Struck and G.\ Burkard, Phys. Rev. B {\bf82}, 125401 (2010).

\bibitem{Heine2013b} P.\ Mir\'{o}, M.\ Ghorbani-Asl and T.\ Heine, Adv. Mater. {\bf25}, 5473-5475 (2013).

\bibitem{Powell64} M.J.D.\ Powell, The Computer Journal.  {\bf7} (2) 155-162. (1964).

\end{thebibliography}
\end{document}